\documentstyle[preprint,tighten,aps,epsfig]{revtex}
\begin{document}
\title{Constituent quark model study of the meson spectra}
\author{J. Vijande, F. Fern\'andez, A. Valcarce}
\address{Nuclear Physics Group,\\
University of Salamanca,\\
Plaza de la Merced s/n, E-37008 Salamanca, Spain}
\maketitle

\begin{abstract}
The $q\bar q$ spectrum is studied in a generalized constituent quark model
constrained in the study of the $NN$ phenomenology and the baryon spectrum. An
overall good fit to the available experimental data is obtained. A detailed
analysis of all sectors from the light-pseudoscalar and vector mesons to
bottomonium is performed paying special attention to the existence and nature
of some non well-established states. These results 
should serve as a complementary tool in distinguishing
conventional quark model mesons from glueballs, hybrids 
or multiquark states.
\end{abstract}

\vspace*{2cm} \noindent Keywords: \newline
nonrelativistic quark models, meson spectrum, scalar mesons
\newline
\newline
\noindent Pacs: \newline
12.39.Jh, 14.40.-n

\newpage

\section{Introduction}

Meson spectroscopy is an extremely broad subject covering from the 
few hundred MeV masses of the light pseudoscalar mesons to the 10 GeV 
scale of the $b\bar b$ system. Such a wide energy region allows to 
address perturbative and nonperturbative phenomena of the underlying 
theory. The continuously increasing huge amount of data and its apparent 
simplicity made mesons ideal systems to learn about the properties of QCD. 
As an exact solution of the theory seems not attainable at present and the big 
effort done in developing its lattice approximation is still not able to 
produce, without guidance, realistic results, phenomenological models are 
the most important theoretical tool to study the meson properties. 
Last years are being extremely exciting due to the new data obtained 
at the $B$ factories, reopening the interest of classifying the meson 
experimental data as $q\bar q$ states according to $SU(N)$ irreducible 
representations. The recently measured $D_{sJ}^*(2317)$ and $X(3872)$ 
states seem to be hardly accommodated in a pure $q\bar q$ description. 
The $D_{sJ}^*(2317)$ presents a mass much lower than the prediction of
potential models, whereas the $X(3872)$ is too light to be a $2P$
charmonium state and too heavy for a $1D$ state. Such discrepancies
could be related with the
structure of some of the light scalar mesons still not well established
theoretically. To be able to understand the nature
of new resonances it is important that we have a template against which to
compare observed states with theoretical predictions.

Much has been learned during the last years 
about the structure and properties of QCD.
The study of charmonium and bottomonium made clear that heavy-quark systems
are properly described by nonrelativistic potential models reflecting
the dynamics expected from QCD \cite{eich}. The {\it a priori} complicated 
light-meson sector was quite surprisingly well reproduced on its bulk 
properties by means of a universal one-gluon exchange plus a linear confining 
potential \cite{godf}. However, the dynamics of the light-quark sector 
is expected to be dominated by the nonperturbative 
spontaneous breaking of chiral symmetry,
basic property of the QCD Lagrangian not satisfied by the
first constituent quark model approaches to the light-meson sector.
Chiral symmetry breaking suggests to divide the quarks into two 
different sectors: light quarks ($u$, $d$, and $s$) where the 
chiral symmetry is spontaneously broken, and heavy quarks ($c$ and $b$) 
where the symmetry is explicitly broken. The origin of the constituent quark 
mass can be traced back to the spontaneous breaking 
of chiral symmetry and consequently constituent quarks
should interact through the exchange of Goldstone bosons \cite{manh},
these exchanges being essential to obtain a correct description of the $NN$
phenomenology and the light baryon spectrum. 
Therefore, for the light sector hadrons can be described as
systems of confined constituent quarks (antiquarks) interacting through
gluons and boson exchanges, whereas for the heavy sector hadrons 
are systems of confined current quarks interacting through gluon exchanges. 
Concerning the other basic property of the theory, confinement,
little analytical progress has been made. Lattice calculations 
indicate that the linear confining potential at short distances 
is screened due to pair creation
making it flat at large distances~\cite{bali}.

A preliminary study of the light-meson spectra and their strong decays
has already been done with an interacting potential constrained on the 
study of the $NN$ system and the nonstrange baryon spectrum \cite{lui}.
Our purpose is to generalize this work obtaining a quark-quark interaction
that allows for a complete study of the meson spectra, from the light
to the heavy sector. To find new physics, it is important that we test
the quark model against known states to understand its strengths and weakness.
To this end we shall begin in the next section discussing the theoretical
ingredients of the constituent quark model. In Sec. III we
perform a detail comparison of the predictions of our model with experiment
from the light pseudoscalar and vector mesons to bottomonium. 
This will allow us to identify discrepancies between the quark model
predictions and experiment that may 
signal physics beyond conventional hadron spectroscopy.
We shall go over these puzzles to decide whether the discrepancy
is most likely a problem with the model or the experiment, or whether it most
likely signals some new physics. In particular the possible presence of
four quark systems in the scalar meson sector will be addressed.  
Finally, in Sec. IV we summarize our most important findings. 

\section{SU(3) constituent quark model}

Since the origin of the quark model hadrons have been considered to be
built by constituent (massive) quarks. Nowadays it is widely
recognized that the constituent quark mass appears because of the spontaneous
breaking of the original $SU(3)_{L}\otimes SU(3)_{R}$ chiral symmetry at
some momentum scale. The picture of the QCD vacuum as a dilute medium of
instantons \cite{diak} explains nicely such a symmetry breaking, which is
the most important nonperturbative phenomenon for hadron structure at low
energies. Quarks interact with fermionic zero modes of the individual
instantons in the medium and therefore the propagator of a light quark gets
modified and quarks acquire a momentum dependent mass which drops to zero
for momenta higher than the inverse of the average instanton size 
$\bar{\rho}$. The momentum dependent quark mass acts as a natural cutoff of the
theory. In the domain of momenta $k<1/\overline{\rho }$, a simple Lagrangian
invariant under the chiral transformation can be derived as \cite{diak}

\begin{equation}
L=\overline{\psi}({\rm i} \gamma^\mu \partial_\mu -MU^{\gamma _{5}})\psi
\label{eq1}
\end{equation}
being $U^{\gamma _{5}}=\exp (i\pi ^{a}\lambda ^{a}\gamma
_{5}/f_{\pi })$. $\pi ^{a}$ denotes the pseudoscalar fields $(\vec{\pi }
,K_{i},\eta_8)$ with i=1,...,4, and $M$ is the constituent quark mass. An
expression of the constituent quark mass can be obtained from the theory,
but it also can be parametrized as $M=m_{q}F(q^{2})$ with
\begin{equation}
F(q^{2})=\left[ \frac{\Lambda^{2}}{\Lambda^{2}+q^{2}} \right] ^{\frac{1}{2}}
\label{eq2}
\end{equation}
where $\Lambda$ determines the scale at which chiral symmetry is
broken. Once a constituent quark mass is generated such particles have
to interact through Goldstone modes. Whereas the Lagrangian
$\overline{\psi} (i\gamma^\mu \partial_\mu -M)\psi$ is not invariant under
chiral rotations, that of Eq. (\ref{eq1})
is invariant since the rotation
of the quark fields can be compensated renaming the bosons fields.
$U^{\gamma _{5}}$ can be expanded in terms of boson fields as,    
\begin{equation}
U^{\gamma _{5}}=1+\frac{{\rm i}}{f_{\pi }}\gamma ^{5}\lambda ^{a}\pi ^{a}-
\frac{1}{2f_{\pi }^{2}}\pi ^{a}\pi ^{a}+...
\label{eq3}
\end{equation}
The first term generates the constituent quark mass and the second one gives
rise to a one-boson exchange interaction between quarks. The main
contribution of the third term comes from the two-pion exchange which can
be simulated by means of a scalar exchange potential. 
Inserting Eqs. (\ref{eq2}) and (\ref{eq3}) in Eq. (\ref{eq1}), one
obtains the simplest Lagrangian invariant under the chiral
transformation $SU(3)_{L}\otimes SU(3)_{R}$ with a scale
dependent constituent quark mass, containing $SU(3)$
scalar and pseudoscalar potentials. The nonrelativistic
reduction of this Lagrangian has been performed for the study of
nuclear forces and will not be repeated here, although the interested
reader can follow the particular steps in several theoretical
works \cite{mosz,mach}. The different terms of the potential
contain central and tensor or central and spin-orbit contributions
that will be grouped for consistency. Therefore, the chiral 
part of the quark-quark interaction can be resumed as follows,
\begin{equation}
V_{qq}(\vec r_{ij}) \, = \,
V_{qq}^C(\vec r_{ij}) \, + \,
V_{qq}^T(\vec r_{ij}) \, + \,
V_{qq}^{SO}(\vec r_{ij}) \, ,
\label{teq}
\end{equation}
where $C$ stands for central, $T$ for tensor, and $SO$ for spin-orbit
potentials. The central part presents four different contributions,
\begin{equation}
V_{qq}^C(\vec r_{ij}) \, = \,
V_{\pi}^C(\vec r_{ij}) \, + \,
V_{\sigma}^C(\vec r_{ij}) \, + \,
V_{K}^{C}(\vec r_{ij}) \, + \,
V_{\eta}^C(\vec r_{ij}) \, ,
\end{equation}
being each interaction given by,
\begin{eqnarray}
V_{\pi}^{C}(\vec{r}_{ij})&=&{\frac{g_{ch}^{2}}{{4\pi }}}{\frac{m_{\pi}^{2}}{{\
12m_{i}m_{j}}}}{\frac{\Lambda _{\pi}^{2}}{{\Lambda _{\pi}^{2}-m_{\pi}^{2}}}}%
m_{\pi}\left[ Y(m_{\pi}\,r_{ij})-{\frac{\Lambda _{\pi}^{3}}{m_{\pi}^{3}}}%
Y(\Lambda _{\pi}\,r_{ij})\right] (\vec{\sigma}_{i}\cdot \vec{\sigma}%
_{j})\sum_{a=1}^{3}{(\lambda _{i}^{a}\cdot \lambda _{j}^{a})}\,, \nonumber \\
V_{\sigma}^{C}(\vec{r}_{ij})&=&-{\frac{g_{ch}^{2}}{{4\pi }}}
{\frac{\Lambda_{\sigma}^{2}%
}{{\ \Lambda _{\sigma}^{2}-m_{\sigma}^{2}}}}m_{\sigma}
\left[ Y(m_{\sigma}\,r_{ij})-{\frac{%
\Lambda _{\sigma}}{m_{\sigma}}} 
Y(\Lambda _{\sigma}\,r_{ij})\right]\, , \label{zp}\\
V_{K}^{C}(\vec{r}_{ij})&=&{\frac{g_{ch}^{2}}{{4\pi }}}{\frac{m_{K}^{2}}{{\
12m_{i}m_{j}}}}{\frac{\Lambda _{K}^{2}}{{\Lambda _{K}^{2}-m_{K}^{2}}}}m_{K}%
\left[ Y(m_{K}\,r_{ij})-{\frac{\Lambda _{K}^{3}}{m_{K}^{3}}}Y(\Lambda
_{K}\,r_{ij})\right] (\vec{\sigma}_{i}\cdot \vec{\sigma}_{j})\sum_{a=4}^{7}{%
(\lambda _{i}^{a}\cdot \lambda _{j}^{a})}\, , \nonumber \\
V_{\eta }^{C}(\vec{r}_{ij})&=&{\frac{g_{ch}^{2}}{{4\pi }}}{\frac{m_{\eta }^{2}%
}{{\ 12m_{i}m_{j}}}}{\frac{\Lambda _{\eta }^{2}}{{\Lambda _{\eta
}^{2}-m_{\eta }^{2}}}}m_{\eta }\left[ Y(m_{\eta }\,r_{ij})-{\frac{\Lambda
_{\eta }^{3}}{m_{\eta }^{3}}}Y(\Lambda _{\eta }\,r_{ij})\right] (\vec{\sigma}%
_{i}\cdot \vec{\sigma}_{j})\left[ cos\theta_P(\lambda _{i}^{8}\cdot
\lambda _{j}^{8})-sin\theta _P\right] \, , \nonumber
\end{eqnarray}
the angle $\theta_P$ appears as a consequence of considering
the physical $\eta$ instead the octet one. 
$g_{ch}=m_{q}/f_{\pi }$, the $\lambda ^{\prime }s$ are the $SU(3)$
flavor Gell-Mann matrices. $m_{i}$ is the quark
mass and $m_{\pi}$, $m_{K}$ and $m_{\eta }$ are the masses of the $SU(3)$
Goldstone bosons, taken to be their experimental values. $m_{\sigma}$ is
determined through the PCAC relation 
$m_{\sigma}^{2}\sim m_{\pi}^{2}+4\,m_{u,d}^{2}$ \cite{scad}. 
Finally, $Y(x)$ is the standard Yukawa function defined by $Y(x)=e^{-x}/x$.

There are three different contributions to the tensor potential,
\begin{equation}
V_{qq}^T(\vec r_{ij}) \, = \,
V_{\pi}^T(\vec r_{ij}) \, + \,
V_{K}^{T}(\vec r_{ij}) \, + \,
V_{\eta}^T(\vec r_{ij}) \, ,
\end{equation}
each term given by,
\begin{eqnarray}
V_{\pi}^{T}(\vec{r}_{ij})&=&{\frac{g_{ch}^{2}}{{4\pi }}}{\frac{m_{\pi}^{2}}{{%
12m_{i}m_{j}}}}{\frac{\Lambda _{\pi}^{2}}{{\Lambda _{\pi}^{2}-m_{\pi}^{2}}}}%
m_{\pi}\left[ H(m_{\pi}\,r_{ij})-{\frac{\Lambda _{\pi}^{3}}{m_{\pi}^{3}}}%
H(\Lambda _{\pi}\,r_{ij})\right] 
S_{ij}\sum_{a=1}^{3}{(\lambda _{i}^{a}\cdot \lambda _{j}^{a})}\, , \nonumber \\
V_{K}^{T}(\vec{r}_{ij})& =&{\frac{g_{ch}^{2}}{{4\pi }}}{\frac{m_{K}^{2}}{{%
12m_{i}m_{j}}}}{\frac{\Lambda _{K}^{2}}{{\Lambda _{K}^{2}-m_{K}^{2}}}}m_{K}%
\left[ H(m_{K}\,r_{ij})-{\frac{\Lambda _{K}^{3}}{m_{K}^{3}}}H(\Lambda
_{K}\,r_{ij})\right] 
S_{ij}\sum_{a=4}^{7}{(\lambda _{i}^{a}\cdot \lambda _{j}^{a})}\, , \\
V_{\eta}^{T}(\vec{r}_{ij})& =&{\frac{g_{ch}^{2}}{{4\pi }}}{\frac{m_{\eta
}^{2}}{{12m_{i}m_{j}}}}{\frac{\Lambda _{\eta }^{2}}{{\Lambda _{\eta
}^{2}-m_{\eta }^{2}}}}m_{\eta }\left[ H(m_{\eta }\,r_{ij})-{\frac{\Lambda
_{\eta }^{3}}{m_{\eta }^{3}}}H(\Lambda _{\eta }\,r_{ij})\right] 
S_{ij}\left[ cos\theta_P(\lambda _{i}^{8}\cdot \lambda
_{j}^{8})-sin\theta_P\right] \, , \nonumber
\end{eqnarray}
being $S_{ij} \, = \, 3 \, ({\vec \sigma}_i \cdot
{\hat r}_{ij}) ({\vec \sigma}_j \cdot  {\hat r}_{ij})
\, - \, {\vec \sigma}_i \cdot {\vec \sigma}_j$
the quark tensor operator and $H(x)=(1+3/x+3/x^2)\,Y(x)$.

Finally, the spin-orbit potential only presents a contribution
coming form the scalar part of the interaction, 
\begin{equation}
V_{qq}^{SO}(\vec{r}_{ij}) = 
V_{\sigma}^{SO}(\vec{r}_{ij}) =-{\frac{g_{ch}^{2}}{{4\pi }}}{\frac{\Lambda
_{\sigma}^{2}}{{\Lambda _{\sigma}^{2}-m_{\sigma}^{2}}}}{\frac{m_{\sigma}^{3}}
{{2m_{i}m_{j}}}}%
\left[ G(m_{\sigma}\,r_{ij})-{\frac{\Lambda_{\sigma}^{3}}{m_{\sigma}^{3}}}
G(\Lambda_{\sigma}\,r_{ij})\right] \vec{L}\cdot \vec{S} 
\end{equation}
where $G(x)=(1+1/x)\,Y(x)/x$. The chiral coupling constant 
$g_{ch}$ is determined from the $\pi NN$ coupling constant through 
\begin{equation}
\frac{g_{ch}^{2}}{4\pi }=\left( \frac{3}{5}\right) ^{2}{\frac{g_{\pi NN}^{2}%
}{{4\pi }}}{\frac{m_{u,d}^{2}}{m_{N}^{2}}}
\end{equation}
what assumes that flavor $SU(3)$ is an exact symmetry
only broken by the different mass of the strange quark. 

This interaction, arising as a consequence of the instanton induced
chiral symmetry breaking, gives rise, among other effects, to 
vector-pseudoscalar meson mass splitting and it also 
generates flavor mixing for the $\eta$ mesons. This is the very
same effect obtained by other instanton induced approaches
as those based on the 't Hooft Lagrangian \cite{toof,resa}. 
In the heavy-quark sector chiral symmetry is explicitly broken and
the Goldstone boson exchange interaction is not active
in such a way that we cannot reproduce the 
hyperfine splittings for heavy mesons. This is a common feature of other
instanton induced interactions \cite{shur} unless a 
phenomelogical parametrization is done \cite{bonnm}.
Beyond the chiral symmetry breaking scale one expects the dynamics
being governed by QCD perturbative effects. They mimic the gluon
fluctuations around the instanton vacuum and are taken into account
through the one-gluon-exchange (OGE) potential. Such a 
potential nicely describes the heavy-meson phenomenology.
Following de R\'{u}jula {\it et al.} \cite{ruju}
the OGE is a standard color Fermi-Breit interaction given by the
Lagrangian,
\begin{equation}
L=i\sqrt{4\pi }\alpha _{s}\overline{\psi }\gamma _{\mu }G^{\mu
}\lambda ^{c}\psi
\end{equation}
where $\lambda^{c}$ are the $SU(3)$ color matrices, $G^{\mu}$ is the
gluon field and $\alpha_s$ is the quark-gluon coupling constant. The 
nonrelativistic reduction of the OGE diagram in QCD for point-like
quarks presents a contact term that, when not treated perturbatively, 
leads to collapse \cite{BHA80}. This is why one
maintains the structure of the OGE, but the $\delta$ function is
regularized in a suitable way. This regularization is justified based on
the finite size of the constituent quarks and should be therefore
flavor dependent \cite{YYYY}. As a consequence, the central part of the
OGE reads,
\begin{equation}
V_{OGE}^{C}(\vec{r}_{ij}) ={\frac{1}{4}}\alpha _{s}\,\vec{\lambda ^{c}}%
_{i}\cdot \vec{\lambda^{c}}_{j}\,\left\{ {\frac{1}{r_{ij}}}-{\frac{1}{%
6m_{i}m_{j}}}\vec{\sigma}_{i}\cdot \vec{\sigma}_{j}
\,{\frac{{e^{-r_{ij}/r_{0}(\mu )}}}{r_{ij}\,
r_0^2(\mu)}}\right\}
\end{equation}
where $r_0(\mu)=\hat r_0/\mu$, scaling with the reduced mass as expected
for a coulombic system. Let us note that the nonrelativistic 
reduction of the one-gluon exchange diagram in QCD yields several
spin-independent contributions \cite{ruju} that have not been
considered. As the OGE is an effective interaction we have
included only the relevant different structures obtained from 
the nonrelativistic reduction and neglected the other terms
that are supposed to be included in the 
fitted parameters \cite{godf}.  

The noncentral terms of the OGE present a similar problem. For point-like
quarks they contain an $1/r^3$ term that in spite of its strength has been 
usually treated perturbatively. Once again the finite size of the constituent 
quarks allows for a regularization and therefore for an exact treatment of
these contributions, obtaining tensor and spin-orbit
potentials of the form,
\begin{eqnarray}
V_{OGE}^{T}(\vec{r}_{ij})& =&-{\frac{1}{16}}{\frac{\alpha_s}{{m_{i}m_{j}}}}
\vec{\lambda}_i^c \cdot \vec{\lambda}_j^c \left[ {1 \over r_{ij}^3} - 
{{e^{-r_{ij}/r_g(\mu)}} \over {r_{ij}}} \left( {1 \over r_{ij}^2} + 
{1 \over {3 r_g^2(\mu)}} + {1 \over {r_{ij} \, r_g(\mu)}} \right) \right] 
S_{ij}  \, , \nonumber \\
V_{OGE}^{SO}(\vec{r}_{ij}) & =& 
-{\frac{1}{16}}{\frac{\alpha_s}{{m_{i}^2m_{j}^2}}}
\vec{\lambda}_i^c \cdot \vec{\lambda}_j^c \left[{1 \over r_{ij}^3} - 
{{e^{-r_{ij}/r_g(\mu)}} \over {r_{ij}^3}} \left( 1 + 
{r_{ij} \over { r_g(\mu)}}  \right) \right] \, \times \, \\
&&\left[ \left( (m_{i}+m_{j})^{2}+2m_{i}m_{j}\right) (\vec{S}_{+}\cdot \vec{L%
})+\left( m_{j}^{2}-m_{i}^{2}\right) 
(\vec{S}_{-}\cdot \vec{L})\right] \, , \nonumber
\end{eqnarray}
where $\vec S_{\pm}=\vec S_i\pm\vec S_j$, and $r_g(\mu)=\hat r_g /\mu$
presents a similar behavior to the scaling of the central term.
The wide energy covered to describe the light, strange and
heavy mesons requires an effective scale-dependent 
strong coupling constant \cite{tita} that 
cannot be obtained from the usual one-loop expression
of the running coupling constant because it diverges
when $Q\rightarrow\Lambda_{QCD}$. The freezing of the
strong coupling constant at low energies
studied in several theoretical approaches \cite{shir,bada2}
has been used in different phenomenological models \cite{pre2}.
The momentum-dependent quark-gluon coupling constant is frozen for each
flavor sector. For this purpose one has to determine 
the typical momentum scale of each flavor sector that, 
as explained in Ref. \cite{halz}, can be assimilated to the
reduced mass of the system. As a consequence,
we use an effective scale-dependent strong
coupling constant given by
\begin{equation}
\alpha_s(\mu)={\alpha_0\over{ln\left({{\mu^2+\mu^2_0}
\over\Lambda_0^2}\right)}},
\label{asf}
\end{equation}
where $\mu$ is the reduced mass of the $q\bar q$ system and $\alpha_0$, 
$\mu_0$ and $\Lambda_0$ are determined as explained in Sec. III.
This equation gives rise to $\alpha_s\sim0.54$ for the light-quark sector,
a value consistent with the one used in the study of the nonstrange 
hadron phenomenology \cite{lui,brun}, and it also has an appropriate 
high $Q^2$ behavior, $\alpha_s\sim0.127$ at the $Z_0$ mass \cite{pre1}.
In Fig. \ref{fig1} we compare our parametrization to the experimental 
data \cite{klut,emem}. We also show for comparison the parametrization 
obtained in Ref. \cite{shir} from an analytical model of QCD.

When Goldstone-boson exchanges are considered together with the OGE,
the possibility of double counting emerges. This question connects
directly with the nature of the pion, studied for a long time
concluding its dual character as $q \bar q$ pair and Goldstone 
boson \cite{weis,thom}. In first approximation, it should be
reasonable to construct a theory in which chiral symmetry
is retained in the Goldstone mode but the internal structure
of the pion is neglected. This would be essentially a long-wave
length approximation \cite{thom}. This may be the reason why
the OPE generates contributions that are not
obtained from the OGE, while the $\rho$ and $\omega$ meson exchanges
give rise to contributions already generated by the OGE \cite{ferp}.
Explicit studies on the literature about the double counting
problem concluded that while the pion can be safely
exchanged together with the gluon, the vector and axial
mesons cannot \cite{yazk}.

Finally, any model imitating QCD should incorporate another 
nonperturbative effect, confinement, that takes into account that the
only observed hadrons are color singlets. It remains an unsolved problem
to derive confinement from QCD in an analytic manner. The only indication
we have on the nature of confinement is through lattice studies,
showing that $q\bar q$ systems are well reproduced at short 
distances by a linear potential. Such potential can be physically
interpreted in a picture in which the quark and the antiquark are linked
with a one-dimensional color flux tube. The spontaneous creation of
light-quark pairs may give rise to a breakup of the color flux tube.
It has been proposed that this translates into a screened 
potential \cite{bali}, in such a way that the potential does not
rise continuously but it saturates at some interquark distance.
Although string breaking has not been definitively confirmed
through lattice calculations \cite{bali2}, a quite rapid
crossover from a linear rising to a flat potential is well
established in SU(2) Yang-Mills theories \cite{este}.
A screened potential simulating these results 
can be written as,
\begin{equation}
V^C_{CON}(\vec{r}_{ij})=\{-a_{c}\,(1-e^{-\mu_c\,r_{ij}})+ \Delta\}(\vec{%
\lambda^c}_{i}\cdot \vec{ \lambda^c}_{j})\,
\end{equation}
where $\Delta$ is a global constant fixing the origin of
energies. At short distances this potential presents 
a linear behavior with an effective confinement strength 
$a=a_c \, \mu_c \, \vec{\lambda^c}_i \cdot \vec{\lambda^c}_j$, while
it becomes constant at large distances. Such screened 
confining potentials provide with an explanation to the
missing state problem in the baryon spectra \cite{miss} and also to the
deviation of the Regge trajectories from the linear behavior
for higher angular momentum states \cite{golr}.

One important question which has not been properly answered
is the covariance property of confinement. While the
spin-orbit splittings in heavy-quark systems
suggest a scalar confining potential \cite{luca}, Ref. \cite
{szcz} showed that the Dirac structure of confinement is of vector nature in
the heavy-quark limit of QCD. On the other hand, a significant mixture
of scalar and vector confinement has been used to explain the decay widths
of $P$-wave $D$ mesons \cite{suga}. Nonetheless,
analytic techniques \cite{bram} and numerical
studies using lattice QCD \cite{bbal} have shown that the confining
forces are spin independent apart from the inevitable spin-orbit
pseudoforce due to the Thomas precession \cite{isg2}.
Therefore, we will consider a confinement spin-orbit contribution
as an arbitrary combination of scalar and vector terms,
\begin{eqnarray}
V_{CON}^{SO}(\vec{r}_{ij}) &=&-(\vec{\lambda}_{i}^{c}\cdot
\vec{\lambda}_{j}^{c})\,{\frac{{a_c\,\mu_c 
\,e^{-\mu_c \,r_{ij}}}}{{4m_{i}^{2}m_{j}^{2}r_{ij}}}}
\left[ \left( (m_{i}^{2}+m_{j}^{2})(1-2\,a_{s}) \right. \right. \nonumber \\
&&\left. \left. +4m_{i}m_{j}(1-a_{s})\right) (\vec{S}_{+}\cdot \vec{L})+ 
(m_j^2-m_i^2) (1-2\,a_s)(\vec{S}_{-}\cdot \vec{L}) \right]
\end{eqnarray}
where $a_s$ would control the ratio between them.

Once perturbative (one-gluon exchange) and nonperturbative (confinement
and chiral symmetry breaking) aspects of QCD have been considered, one
ends up with a quark-quark interaction of the form
(from now on we will refer to a light quark,
$u$ or $d$, as $n$, $s$ will be used for the strange quark and $Q$ for the
heavy quarks $c$ and $b$):
\begin{equation}
V_{q_iq_j}=\left\{ 
\begin{array}{ll}
q_iq_j=nn\Rightarrow V_{CON}+V_{OGE}+V_{\pi}+V_{\sigma}+V_{\eta } &  \\ 
q_iq_j=ns\Rightarrow V_{CON}+V_{OGE}+V_{\sigma}+V_{K}+V_{\eta } &  \\ 
q_iq_j=ss\Rightarrow V_{CON}+V_{OGE}+V_{\sigma}+V_{\eta } &  \\ 
q_iq_j=nQ\Rightarrow V_{CON}+V_{OGE} &  \\ 
q_iq_j=QQ\Rightarrow V_{CON}+V_{OGE} & 
\end{array}
\right.  \label{pot}
\end{equation}
The corresponding $q \bar{q}$ potential is obtained from the
$qq$ one as detailed in Ref. \cite{bumu}. In the case of $V_K(\vec{r}_{ij})$,
where $G$ parity is not well defined, the transformation
is given by $\lambda^a_1 \cdot \lambda^a_2 \to
\lambda^a_1 \cdot (\lambda^a_2)^T$, which recovers the standard
change of sign in the case of the pseudoscalar exchange between
two nonstrange quarks.

Apart from models incorporating the one-gluon exchange potential 
for the short-range part of the interaction, one finds in the 
literature other attempts to study hadron phenomenology based on instanton
induced forces. In Ref. \cite{okann} a flavor antisymmetric quark-quark 
instanton induced interaction was derived. It was used in
a nonrelativistic framework for the
study of the two-baryon system and the baryon spectrum, but it was never 
applied to study the meson spectrum. Such a model does
not contain Goldstone boson exchanges, which are essential to
make contact with the one and two baryon systems. The short-ranged instanton 
induced force was supplemented by a baryonic meson exchange potential 
to give a quantitative explanation of experimental data. 
Moreover, in Ref. \cite{shuro} it was demonstrated that this 
instanton induced force reproduces the baryon 
spectrum as well as the one-gluon exchange. The other 
extensive work using instanton induced interactions
is that of the group of Bonn \cite{bonnz,bonnep}. The interaction
is derived in Ref. \cite{bonnz} obtaining a sum of contact
terms that are regularized (as we have done for the OGE
potential) and whose flavor mixing matrix elements
are fitted to experimental data.
This interaction was supplemented by a phenomenological confining
potential and applied in a Bethe-Salpeter framework to study
the light meson and baryon spectra. This 
instanton induced force is only valid for light quarks in such a
way that the extension of the model to study heavy flavors 
is done in a completely phenomenological way \cite{bonnm}.
Therefore, these instanton induced models do not allow for the moment
for a coherent study of the light and heavy flavors neither for a simultaneous 
description of the baryon-baryon phenomenology.
These two approaches, the one based on the 't Hooft interaction and
the one followed in the present work, clearly shares the most important
features of the quark-quark interaction. The reason why we adopt
the first scheme lies in the fact that it includes explicitly 
Goldstone boson exchanges between quarks which are essential to make
contact with the one and two-baryon phenomenology.

\section{Results and discussion}

Let us first discuss how the parameters of the model are fixed.
Most of them, as for example those of the Goldstone boson fields, are 
taken from calculations on two-baryon systems.
Once the Goldstone boson exchange part
of the interaction has been determined, the one-gluon exchange
controls the hyperfine splittings. It is worth to notice
the relevance of the OGE contribution for the description
of the meson spectra, as has already been emphasized in Ref. \cite{lui}.
The conclusion of this work, that the $\rho-\pi$ mass difference
cannot be obtained from the Goldstone-boson exchanges alone
is fully maintained. For the sake of completeness let us 
mentioned that in the present model 
this mass difference would be around 38 MeV if 
the OGE is not used. As it is also illustrated in Ref. \cite{lui}
the dependence of the $\rho - \pi$ mass difference on the 
Goldstone-boson cut-off masses is small, in such a way
that the spectra come never determined by the election
of the cut-off masses.
We have fixed the OGE parameters by 
a global fit to the hyperfine splittings well established by
the Particle Data Group (PDG) \cite{pdgb} from the light to the 
heavy-quark sector. The strength of confinement determines
the energy difference between any $J^{PC}$ meson ground state and its 
radial excitations. We have fitted $a_c$
and $\mu_c$ in order to reproduce two well measured energy differences:
the $\rho $ meson and its first radial excitation and the $J/\psi$ 
and the $\psi (2S)$, obtaining values close to those 
inferred in Ref. \cite{pedr} through the analysis of the screening
of QCD suggested by unquenched lattice calculations in the 
heavy-quark sector. Finally, it only remains to fix the
relative strength of scalar and vector confinement. For this purpose one 
has to look for some states where the spin-orbit contribution,
being important, can be easily isolated from other effects
as for example flavor mixing. This is the case of the isovector mesons
$a_{1}(1260)$ and $a_{2}(1320)$. In Fig. \ref{fig2} we
have plotted their masses as a function of $a_s$, the relative
strength of scalar and vector confinement. As can be seen,
using a strict scalar confining potential, $a_s = 1$, one would obtain 
1343 MeV and 1201 MeV, respectively, in
complete disagreement with the ordering and magnitude of
the experimental data. Introducing a mixture
of vector confinement, $a_s$=0.777, the experimental
order is recovered being now the masses 1205 MeV and 1327 MeV, respectively,
both within the experimental error bars. This value also allows to obtain
a good description of the experimental data in the $c\bar c$ and
$b\bar b$ systems. The parameters of the model are resumed in Table \ref{t1}.

Acceptable results for the meson spectra
have been provided by relativistic as well as 
nonrelativistic approaches \cite{lucp}. 
In both cases a QCD-inspired interaction
is used, the difference being in quarks masses and kinematics.
In fact, several works in the literature \cite{sema}
have studied the connections existing between relativistic,
semirelativistic, and nonrelativistic potential models of quarkonium
using an interaction composed of an attractive Coulomb potential
and a confining power-law term. The spectra of
these very different models become nearly similar provided
specific relations exist between the dimensionless parameters
peculiar to each model.

As a consequence we will solve the Schr\"odinger equation for the
relative motion of the $q \bar q$ pair with the interacting 
potential of Eq. (\ref{pot}). There is a particularly simple and 
efficient method for integrating this type of second order differential 
equations, the commonly called Numerov method \cite{koon}.
The noncentral potentials (tensor and spin-orbit) give their
most important contribution for the diagonal terms.
For example for the isovector states
$J^{PC}=1^{--}$ ($L=0$ or $2$, $S=1$, $J=1$) we would have 
the following matrix elements (in MeV):
\begin{equation}
\left( \matrix{
<^3S_1|H|^3S_1>=772 &  <^3S_1|H|^3D_1>=22 \cr
<^3D_1|H|^3S_1>=22  &  <^3D_1|H|^3D_1>=1518} \right) \, ,
\end{equation}
and for the isovector $J^{PC}=2^{++}$ ($L=1,3$, $S=1$, $J=2$)
\begin{equation}
\left( \matrix{
<^3P_2|H|^3P_2>=1327 &  <^3P_2|H|^3F_2>=5 \cr
<^3F_2|H|^3P_2>=5   &  <^3F_2|H|^3F_2>=1797} \right) \, ,
\end{equation}
where the matrix elements are calculated with the wave functions
solution of the single-channel Schr\"odinger equation for the 
total hamiltonian, including central and noncentral terms.
The perturbative effect of the nondiagonal contributions of the
noncentral potentials is observed.
The importance of the diagonal contributions can be easily inferred from the
energies obtained when the different interactions are connected.
In table \ref{tnw} we give the energies obtained when solving
the Schr\"odinger equation for two different partial waves: one
of the cases discussed above, the isovector $^3D_1$, and the
isoscalar $^3P_0$ $(n\overline{n})$. 
One can see that the corrections induced by the diagonal contributions of
the noncentral potentials are much important than the nondiagonal ones.
This effect becomes clearly nonperturbative for the isoscalar
$^3P_0$ state, that on the other hand does not admit coupling to
partial waves with different orbital angular momentum. Therefore
the noncentral potentials will be treated exactly in their
diagonal contributions and by diagonalizing the corresponding matrix
when nondiagonal contributions are present.
The same reasoning as before applies in the case of the spin-orbit
contribution proportional to $S_-$ as will be discussed in Sect.
\ref{3h}.

In Tables \ref{t2} to \ref{t8} we compare the masses obtained
within the present model to experimental data \cite{pdgb}. 
Tables \ref{t2}, \ref{t2b}, and \ref{t3} do not include 
the scalar mesons which will be discussed separately in Sec. \ref{scalar}
and are shown in Table \ref{t8}. In all cases we show the 
name and mass of the experimental
state and the prediction of our model together with the $J^{PC}$ quantum 
numbers, or $J^P$ if $C$ parity is not well defined (in these cases
we have explicitly indicated the spin). If there is an experimental 
meson without assignment of quantum numbers, those indicated
by question marks, and we find a
corresponding state in our model, we indicate 
between square brackets its quantum numbers. 
In the tables we also include in parenthesis the
radial excitation and the orbital angular momentum corresponding
to the state under consideration.

A final comment about the flavor mixing angles is
in order. Our interacting hamiltonian needs as input the pseudoscalar
octet-singlet mixing angle (see Eq. (\ref{zp})). For this angle, values in the
range of $-10^o$ to $-23^o$ have been obtained depending on the 
analysis performed \cite{pdgb,kloe,cbcb,gilm,feld,brae,bura}. 
We have taken an intermediate value of $-$15$^o$.
The model provides a theoretical mixing $n\bar n \leftrightarrow s\bar s$ 
($\omega \leftrightarrow \phi$, $\eta \leftrightarrow \eta'$, ...) 
as a consequence of the $V_K$ potential. The mixing angle obtained is 
$\theta_V=$ 34.7$^o$ for the $\omega \leftrightarrow \phi$ case, 
and $\theta_P= -21.7^o$ for the $\eta \leftrightarrow \eta'$. We want
to emphasize the independence of our result on the input value for 
$\theta_P$. For example if we had used a theoretical input 
of $\theta_P=-23^o$, our predicted
pseudoscalar mixing angle would have been $\theta_P=-21.9^o$, 
almost the same as before. Our result is compatible with many
others reported in the literature \cite{cbcb,feld,bura} 
and close to the $\theta_P^{lin}$ value given on the PDG 
\cite{pdgb} ($\theta_P^{lin}=-23^o$), as expected because we use a mixing
formula based on linear and not quadratic masses.
The mixing angle obtained for most meson nonets is
approximately ideal, exceptions are the nonet of 
pseudoscalar and scalar mesons.
The mixing angle, calculated 
for the ground state of each $J^{PC}$ nonet, is assumed 
for the radial excitations with the same quantum numbers. 

There are a number of states whose quantum numbers
are not clearly determined or do not seem to present a clear correspondence
with a $q\bar q$ state. Let us analyze in detail the results.

\subsection{$I=0$ $J^{PC}=0^{-+}$ states}

There is an overall good agreement between the constituent quark
model results and the experimentally observed mesons.
Recently there has been a modification in the experimental
situation. In the 2002 PDG \cite{PDG02b} there were two
resonances in the 1.5 GeV energy region, $\eta(1295)$ and $\eta(1440)$.
However, the 2004 PDG reports three states, $\eta(1295)$,
$\eta(1405)$ and $\eta(1475)$, all of them interpreted as pseudoscalar $0^{-+}$
states. This modification was mainly due to a recent experiment by the E852
Collaboration~\cite{e852}. They found evidence for three pseudoscalar resonances: $\eta(1295)$,
$\eta(1416)$ and $\eta(1485)$ in the analysis of the reaction $\pi^- p\to K^+K^-\pi^0 n$.
The first one decays exclusively into $a_0
\pi^0$, the second into $a_0 \pi^0$ and $K^* \bar{K}$ and the last one
exclusively into $K^{\ast}\bar{K}$.
We only obtain two $0^{-+}$ states in this energy region, one at 1290 MeV
and the other at 1563 MeV, both being predominantly $n\bar n$.
Based on our results one would identify the first state with the
$\eta(1295)$. However, our model predicts only one state
corresponding either to $\eta(1405)$ or $\eta(1475)$.
Although the energy seems to favor an assignment of our second state
with the $\eta(1475)$, its dominant $n\bar n$ component makes difficult to
explain its experimental decay modes ($K^*\bar K$). However, the
expectation of the $\eta(1475)$ being a dominantly $s \bar s$ $2 \, ^1S_0$ (sometimes
the discussion is more clear using the spectroscopic notation $n \, ^{2S+1}L_J$)
state \cite{ANIS} as suggested by the observation of a large $K^* \bar K$
decay mode is highly dependent on the strong $KK$ final state interaction \cite{brna}. 
The existence of the third resonance would therefore imply the presence of additional states
beyond the two obtained in the quark model. However the experimental situation
is not definitively settled. There are some speculations that one of these states could
be somehow related to the $f_1(1420)$ \cite{napo}. To disentangle the flavor content of the $\eta(1405)$
and $\eta(1475)$ could be a very important experimental contribution feasible at CLEO, what
would help to discriminate among different theoretical models.
  
To illustrate the uncertain situation with these resonances let us finally
mention that the analysis of the $K\bar K\pi$ and $\eta\pi\pi$ channels in
$\gamma\gamma$ collision performed in Ref.~\cite{Ac00m} observed
the $\eta(1475)$ in $K\bar K\pi$, but not the $\eta(1405)$ in $\eta\pi\pi$. Since
gluonium production is presumably suppressed in $\gamma\gamma$ collisions
this result suggests that the $\eta(1405)$ may have a large glueball component
\cite{LY03}.  This interpretation, however, is not favored by lattice calculations, 
which predict the $0^{-+}$ state above 2.5 GeV. 

\subsection{$I=0$ $J^{PC}=1^{++}$ states}

The historically confused experimental status of light axial vectors has 
improved a lot with high statistic central production experiments on 
$\eta\pi\pi$, $K\,K\pi$ and $4\pi$ by WA102 \cite{bar1} and $K\,K\pi$ by E690
\cite{sosa}. In these experiments very clearly $f_1(1285)$ and $f_1(1420)$
states have been observed, but there is no evidence of $f_1(1510)$. Although
the $f_1(1420)$ and the $f_1(1510)$ are well separated in mass and resolved
in different experiments, there are no experiment or production reactions in
which both states have been detected. All these observations express
skepticism regarding the existence of the $f_1(1510)$ \cite{clos}. In view
of their masses the obvious assumption is that the $f_1(1285)$ is the
$1\,^3P_1$ $n\bar n$ state and the $f_1(1420)$ its $s\bar s$ 
partner. In Ref. \cite{clos} a significant
singlet-octet mixing for the nonet of the $f_1(1285)$ and $f_1(1420)$ has 
been obtained from several independent analysis, $\theta\sim50^o$. 
We obtain a value of $46.3^o$ in complete 
agreement with the former study. One should remember that the mixing in
our model is fully determined by the structure of the potential.

\subsection{$I=0$ $J^{PC}=2^{-+}$ states}

Regarding the $I=0$ $2^{-+}$ the situation seems to be clear. Experimentally
there are two states: the $\eta _2(1645)$ with a mass of 1617$\pm$5 MeV 
and decaying into $a_2(1320)\pi$, and the $\eta _2(1870)$ with a mass of 
1842$\pm$8 and decaying into $a_2(1320)\pi$ and $4\pi$. 
We obtain three theoretical 
states in this energy region: the first one with a mass of 1600 MeV and
pure light-quark content, a second with a mass of 1853 MeV and 
pure $s\bar s$ content and a third state with a mass of 1863 MeV and 
also pure light-quark content. 
The assignment of our first state to the $\eta_2(1645)$ seems clear. With 
respect to the $\eta_2(1870)$ its reported decay modes are not accessible in
the $^3P_0$ model having a pure $s\bar s$ content. This enforces to 
assign the $\eta_2(1870)$ to the second excited state with light-quark
content. Thus, it may exist an unobserved resonance close
to the $\eta_2(1870)$ being a pure $s \bar s$ state. 

\subsection{$I=0$ $J^{PC}=1^{--}$ and $J^{PC}=3^{--}$ states}

In the PDG there are three isoscalar $1^{--}$ states in the 1.5 GeV energy
region: $\omega(1420)$, $\omega(1650)$ and $\phi(1680)$. The first and third
state seem to be fairly well established, however in the last two years
there have been several modifications on the mass of the $\omega(1650)$. The
2002 PDG gave a mass of $1649\pm24$ MeV, the 2003 electronic
version reported a mass between $1600-1800$ MeV, while the 2004 PDG
quotes a value of 1670$\pm$30 MeV. This modification is based on the analysis of the reaction
$e^+e^-\rightarrow \pi^+\pi^-\pi^0$~\cite{acha} where two $\omega$ states are
reported, the first one with a mass of $1490\pm75$ MeV decaying into $\pi^+\pi^-\pi^0$
and the second with a mass of $1790\pm 50$ and decaying, with approximately
the same probability, into 3$\pi$ and $\omega\pi\pi$. While
the first state clearly corresponds to the $\omega(1420)$, the
second state was included by the electronic 2003 PDG as an experimental result for
the $\omega(1650)$, increasing drastically the error bars. In the 2004 PDG this experiment has not
been considered within the statistical fit, therefore reducing again the error bars.
A similar situation may happen with the $\phi(1680)$.
The FOCUS Collaboration at Fermilab \cite{link} has reported a high
statistic study of the diffractive photoproduction of $K^+K^-$
confirming the existence of a clear enhancement of the cross section
corresponding with a fitted mass of $1753\pm 4$ MeV.
 
Theoretically we find three $1^{--}$ states in this energy region: two
almost pure light-quark states with masses
of 1444 MeV and 1784 MeV, and an almost pure $s\bar s$ state with a mass of
1726 MeV. Based on the aforementioned discussion, the first two states
may correspond to the $\omega(1420)$ and $\omega(1650)$,
whereas the third one would correspond to the $\phi(1680)$.
Our analysis also assigns the $\omega(1650)$ to
a $3 \, ^3S_1$ state instead of a $1 \, ^3D_1$ wave \cite{ANIS},
that would have a mass of 1475 MeV with a pure
light-quark content, in complete disagreement with the experimental data.   

For the $3^{--}$ case we obtain two almost 
degenerate states with completely different
flavor content. Experimentally there is a clear
evidence that the $\phi_3(1850)$ has a dominant $s\bar s$ flavor content,
and therefore it would correspond to our $s \bar s$ state
at 1875 MeV, predicting the existence of an unobserved $n\bar n$ $3^{--}$ 
state in the same energy region.

\subsection{$I=0$ $J^{PC}=2^{++}$ states}

Experimentally there is a proliferation of 
isoscalar $2^{++}$ states in an energy region
that has been suggested as coexisting with $2^{++}$ glueballs. Theoretically
the results of our model may confirm that the $f_2(1270)$ and $f_2'(1525)$ are
the $n\bar n$ and $s\bar s$ members of the $1\,^3P_2$ $q\bar q$ flavor nonet.
Although $n\bar n \leftrightarrow s\bar s$ mixing is present due to the
kaon exchange, the $n\bar n$ content of the $f_2'(1525)$ is smaller than 
$0.1\%$ in agreement with the experimental $f_2'(1525)\gamma\gamma$ coupling,
which limits the $n\bar n$ content to a few percent \cite{brna}. 
Our model also finds that the $f_2(1565)$, observed in 
$p\bar p$ annihilation at rest, and the $f_2(1640)$, decaying into 
$\omega\omega$ and $4\pi$, could be the same state, the $n\bar n$ 
member of the $2\,^3P_2$ $q\bar q$ flavor nonet as is suggested in the PDG.
The $f_2(1950)$ would be its $s\bar s$ partner in agreement with the 
experimentally observed decays.
Finally the $f_2(1810)$ corresponds to the $n\bar n$ member of the $1\,^3F_2$
nonet and the $f_2(1910)$ to the $n\bar n$ member of the $3\,^3P_2$
nonet, as expected from the decay patterns. 

In this sector there is a meson, the $f_2(1430)$, that has no 
equivalent in our $q\bar q$ scheme. This state is not
confirmed in the PDG and even recent measurements have suggested a 
different assignment of quantum numbers, being a $0^{++}$ state \cite{vlad},
what could make it compatible with the lightest scalar glueball 
\cite{morn}. It was seen together with another resonance, the
$f_2(1480)$, in an experiment where the $f_2(1270)$ was not detected 
\cite{ache}. A full understanding of the nature of the $f_2(1430)$ 
will probably also require an explanation for the absence of the 
$f_2(1270)$ in the experimental data of Ref. \cite{ache}. 

\subsection{$I=0$ $J^{PC}=1^{+-}$ states}

While the $n \bar n$ $1\, ^3P_1$ quark model prediction, 1257 MeV, 
differs slightly from 
its corresponding experimental state, $f_1(1285)$, 
the $n \bar n$ $1\, ^1P_1$ is a little bit far
from the expected corresponding experimental state, $h_1(1170)$. 
Annihilation contributions
could improve the agreement with data, having in mind that
they are expected to be negative and larger in the $S=0$
light-quark sector while almost negligible for $S=1$ states \cite{godf}.
There are some estimations of the contribution of annihilation in the 
light-quark sector, but its dependence on fitted parameters
prevents from making any definitive conclusion.
The $h_1(1380)$ it is a convincing candidate for the $s\bar s$ partner of
the $1\,^1P_1$ $h_1(1170)$. Although a 
first glance to the quark model result seems
to indicate that the mass is too high, the most recent
experimental measurement for this state reports a mass of 1440$\pm60$ 
\cite{h113}. The last measured $1^{+-}$ state, 
the $h_1(1595)$, will correspond to the $n\bar n$ $2 \, ^1P_1$ state, 
its $s\bar s$ partner being an unobserved meson around 1973 MeV 
as also noticed from the analysis of strange decays in Ref. \cite{brna}.

\subsection{$I=1$ mesons}

The results shown in Table \ref{t2b} present an almost perfect
parallelism between the theoretical predictions and
the experimentally observed states. The only not well defined
correspondence comes from the
$\rho(1700)$, which is compatible with an $S$ or $D$ wave.
Our calculation indicates a mass for this state a little bit higher than
the one reported in the PDG and compatible with the observations of 
Ref. \cite{ABEL}.

\subsection{$I=1/2$ strange mesons}
\label{3h}

From a theoretical perspective strange mesons exhibit explicit flavor
and therefore they do not present the additional complications of 
annihilation mixing what makes the isoscalar mesons a priori much more 
complicated.

There is a good correspondence between the $0^-$ and $1^-$ quark model
results and the experimental mesons. In particular, the $K(1630)$, 
whose quantum numbers have not been yet determined, would correspond to the
$2\,^3S_1$ $n\bar s$ state. The only problem in this sector
appears for the $K^*(1410)$, in clear disagreement with the predictions
of the quark model. The $K^*(1410)$ could be the radial excitation 
of the $K^*(892)$, because it is the first $1^{-}$ vector resonance observed.
However, the first $n\bar s$ radial excitation is predicted at 1620 MeV. 
This is reasonable taking into account that in the $S=1$ light-quark sector 
the radial excitation has a mass around 700 MeV above the ground state, what
confirms that the first radial excitation of the $K^*(892)$ 
should be around 1600 MeV. This is why we assign in 
Table \ref{t3} our 1620 MeV state to the $K(1630)$. It would be interesting for
future experiments to check the quantum numbers of this state.
With respect to the $K^*(1410)$ its assignment to the $2\,^3S_1$ state
is not only excluded by its mass, but also by its decay modes. A possible
interpretation of the low mass of this state due to the mixing with hybrids
has been suggested in Ref. \cite{brna} and therefore seems to be excluded as a
pure $q\bar q$ pair.

The $2^+$ states are also in reasonable agreement with the theoretical
predictions. This can be attributed to the understanding of the 
$(1P)\,2^{++}$ and $(1F)\,2^{++}$ isoscalar light-quark sector. The only 
difference is the mass of the strange quark in the interacting potential
what makes clear the $n\bar s$ structure of $K_2^*(1430)$ and $K_2^*(1980)$.

Strange mesons are the lighter ones where a mixing of the 
$1^+$ states with $S=0$ and $S=1$ can occur (also the $2^-$ states) due to the
fact that $C$ parity is not a good quantum number. Although there is not
theoretical consensus about the origin of such a mixing, in our model it is
induced by the spin-orbit contributions of the OGE and confinement potentials
proportional to $\vec S_-$. As a consequence the physical states would be
given by
\begin{eqnarray}
\mid K_1^* \rangle &=&\cos \theta \mid 1^+(S=0)\rangle-\sin \theta 
\mid 1^+(S=1)\rangle \nonumber \\
\mid K_1 \rangle &=&\sin \theta \mid 1^+(S=0)\rangle+\cos \theta 
\mid 1^+(S=1)\rangle
\label{mix}
\end{eqnarray}
In the literature this mixing angle has been estimated by 
different methods. In Ref. \cite{rosn} it was calculated from the ratio 
$B(\tau\to\nu K_1(1270))/B(\tau \to \nu K_1(1400))$, obtaining a value
$\theta\sim+62^o$. In Ref. \cite{brna} an angle $\theta\sim+45^o$ 
was calculated from the pattern of the decay branching
fractions of the two experimental states, $K_1(1270)$ and $K_1(1400)$.
HQET \cite{iswi} gives two possible mixing
angles, $\theta\sim+35.3^o$ and $\theta\sim-54.7^o$. We obtain a mixing
angle of $\theta\sim55.7^o$ and therefore a mass for the physical states of
1352 MeV and 1414 MeV, improving the agreement with the experiment with 
respect to the unmixed masses. The mixing angle is close to the
preferred value by the experimental decays and far from the results
of Ref. \cite{koko} in a relativized quark model based only on OGE and
confinement potentials. Their mixing angle $\theta\sim+5^o$ seems to be
definitively excluded by the experimental decays.

The same mixing occurs for the $(2P)\,1^+$ 
states, giving two resonances around 1.85 GeV, one would
correspond to the $K_1(1650)$ and the other to an unobserved state. 
Although this identification could seem inconsistent, the PDG
states that the $K_1(1650)$ entry contains 
various peaks reported in partial-wave
analysis in the 1600$-$1900 mass region. The election of a mass of 1650 MeV
is done based on three measurements, one of $1650\pm50$ MeV and two around 
1840 MeV. The same identification for $K_1(1650)$ together with the 
prediction of a $2P$ tensor state $K_2^*(1850)$ has been done in 
Ref. \cite{brna} from the analysis of experimental data on strong decays.

We have included in Table \ref{t3} a state given in the PDG as $K_2(1580)$
that has no equivalent in the $q\bar q$ spectrum. This state is 
clearly uncertain, it was reported in only one experimental work more
than twenty years ago and has never been measured again.

\subsection{$D$ and $D_s$ mesons}

Recently, a clear evidence for the existence of new open-charm mesonic states 
has been reported by three different collaborations, two states
with charm-strange quark content and other with charm-light quark content.

BaBar Collaboration 
reported a narrow state at 2316.8$\pm$0.5 MeV \cite{baba} 
called $D_{sJ}^*(2317)$. CLEO Collaboration \cite{cleo,cleo3,cleo5} provided
confirmation of the existence of this state and furthermore reported the
observation of a new state called $D_{sJ}(2463)$. 
In return, BaBar experiment
also confirmed the existence of this state \cite{bab2}. Finally both 
discoveries have been confirmed by the Belle Collaboration, not only in the
analysis of inclusive $e^+e^-$ annihilation \cite{bel3}, but also in the 
exclusive B meson decays \cite{bel4}. All the experimental observations
are consistent with the assignment of $P$-wave states with spin-parity
$J^P=0^+$ for the $D_{sJ}^*(2317)$ and $J^P=1^+$ for the $D_{sJ}(2463)$.
Belle Collaboration \cite{bel5} has also reported the existence of a $J^P=0^+$
charm-light state with a mass of 2308$\pm$36 MeV.

There are two intriguing aspects of the new $D$ mesons. 
First of all they have
a mass significantly smaller than the predictions of most QCD inspired 
quark potential models regarding these mesons as $P$-wave bound states,
$c\bar s$ and $c\bar n$ respectively. Secondly they do not show the 
decay patterns favored by theoretical expectations. 
This has triggered a variety of articles either 
supporting the $q\bar q$ interpretation or presenting alternative hypothesis 
of exotic states.

Our results are shown in Tables \ref{t4} and \ref{t9}. In Table \ref{t4} we
have only included those states 
reported by the PDG, while in Table \ref{t9} we include
all $1P$ excited states. As can be seen the low-lying
$0^-$ and $1^-$ $D$ and $D_s$ states are perfectly reproduced.
We obtain a theoretical state with the mass of the $D^*(2640)$ and 
spin-parity $J^P=0^-$, and also a state with a similar mass to the 
$D_{sJ}(2573)$ and $J^P=2^+$. However there is a 
discrepancy both with the $J^P=1^+$ states and specially with the 
$J^P=0^+$ states. In the case of
the $1^+$ states the same mixing as discussed in Eq. (\ref{mix}) between the 
$^1P_1$ and $^3P_1$ partial waves appears. 
As a consequence, for the charm-light sector we obtain
two $1^+$ states with masses of 2454 MeV and 2535 MeV, with a mixing angle
of 43.5$^o$, and for the charm-strange case 2543 MeV and 2571 MeV,
with a mixing angle of 58.4$^o$. Therefore, while we would find a candidate
for the $D_{s1}(2536)$ and the $D_1(2420)$, the recently 
measured $D_{sJ}(2463)$ and its equivalent state in the charm-light
sector (see Table \ref{t9}) do not fit into the predictions of the 
nonrelativistic $q\bar q$ models \cite{luch}.

The situation with the $0^+$ states being in principle more discouraging is,
however, similar to the problem observed for the light-scalar mesons
(see Sec. \ref{scalar}), they hardly fit in a $q\bar q$ scheme. As seen 
in Table \ref{t9} the quark model predictions for the $0^{+}$ $D$ and 
$D_s$ are around 150 MeV above the experimental results. 
In a pure $q\bar q$ scheme the $0^+$ states are influenced by the noncentral
terms of the interaction, in particular they strongly depend on the 
spin-orbit force and therefore on the scalar/vector rate in the potential, 
controlled by $a_s$.  As this relation was fixed in the light-meson sector, 
where the potential has a much more involved structure (Goldstone boson 
exchanges), we wonder if the disagreement
could be solved by modifying the scalar/vector rate of confinement. In 
Table \ref{t10} we present the results obtained by fixing $a_s$ to 
reproduce the experimental mass of one of the new measured $D_s$ mesons, the
$D^*_{sJ}(2317)$. Apart from obtaining a completely unusual value for 
the scalar/vector rate, $a_s=0.46$ \cite{szcz,suga}, in doing this 
one observes how the situation of the charm-light sector does not improve,
what makes evident the rather complicated situation appeared with
the new measurements.

Let us finally note that in a $q \bar q$ scheme the $0^+$ states are 
obtained through an orbital angular momentum excitation, what could explain the large masses
predicted. The $0^+$ quantum numbers may also be obtained in the absence
of orbital excitation considering more complex structures \cite{JAFF} which can
help to understand the experimental situation. 

In the literature the new $(0^+,1^+)$ states have triggered other alternative
explanations as for example that they could be the members of the
$J^P=(0^+,1^+)$ doublet predicted by the heavy quark effective theory (HQET)
\cite{dean} or that these states together with the lowest $(0^-,1^-)$
constitute the chiral doublet of the $(0^+,1^+)$ HQET spin multiplet \cite{bard}.

\subsection{Charmonium and the new states}

We present in Table \ref{t5} our results for charmonium. 
Confinement parameters were fitted to describe the energy difference
between the $S=1$ ground state, $J/\psi(1S)$, and its first radial excitation,
$\psi(2S)$. One also observes a pretty good description 
of the first negative parity orbital excitation, $\psi(3770)$. 
The splitting among the $\chi_{cJ}$ states is 
correctly given both in order and magnitude. With 
respect to the $S=0$ sector, as happened for the light-quark systems, the
orbital excitation is expected to be 600 MeV above the ground state, in
complete agreement with the mass of the $h_c(1P)$. This state, although not
yet confirmed is a clear candidate for being the 
$1^{+-}$ resonance. There is also a set of $L=$even $S=1$ states with 
an obvious correspondence to $q\bar q$ states although in some cases the
$S$ or $D-$wave identification is not unique, 
as seen on Table \ref{t5} for the $\psi(4415)$.

Recently the Belle Collaboration has reported the observation of a narrow
peak that has been interpreted as the $2S$ singlet charmonium state, 
the $\eta_c(2S)$,
\[
M[\eta_c(2S)]=\left\{
\begin{array}{c}
3654 \pm 10 \text{ MeV Ref. \cite{bel1}} \\
3622 \pm 8 \text{ MeV Ref. \cite{bel4}} \\
3630 \pm 8 \text{ MeV Ref. \cite{bel3}}
\end{array} \right. \, .
\]
These masses are larger than the experimental value quoted by the 2002 PDG:
$M[\eta_c(2S)]=3594 \pm 5$ MeV and it was pointed out that cannot 
be easily explained in the framework of constituent quark models.
The reason for that stems on the
$2S$ hyperfine splitting (HFS) that would be smaller than the predicted 
for the $1S$ ones. In fact the theoretical predicted $2S/1S$ HFS ratio is
\[
R=\frac{\Delta M_{2S}^{HFS}}{\Delta M_{1S}^{HFS}}= \left\{
\begin{array}{c}
0.84 \text{ Ref. \cite{eber}} \\
0.67 \text{ Ref. \cite{hfs2}} \\
0.60 \text{ Ref. \cite{hfs3}}
\end{array} \right. \, ,
\]                              
larger than the experimental value, R=0.273 if $M[\eta_c(2S)]=$ 3654 MeV,
R=0.547 for 3622 MeV, and R=0.479 for 3630 MeV.
Our result is $M[\eta _c(2S)]=$ 3627 MeV, within
the error bar of the last two Belle measurements, the ones
obtained with higher statistics. Moreover the ratio $2S/1S$ HFS is
found to be 0.537, in agreement with these last two experimental data. The
reason for this agreement can be found in the radial structure of the
confining potential that also influences the HFS, the
linear confinement being not enough flexible to
accommodate both excitations \cite{pedr}. A similar agreement has also been
obtained by other theoretical models including the effect of the open
thresholds \cite{eic1} (note that the result of 
Ref. \cite{godf} was far from the available experimental data at that time).

Finally HFS is closely connected with the leptonic decay widths 
$V({\rm vector \,\, meson})\to e^+e^-$. Although their absolute values
depend on radiative and relativistic corrections, the ratios are a test
of the wave functions at the origin, and still closely connected to HFS.
We obtain 
$\Gamma_{e^+e^-}[\psi(2S)]/\Gamma_{e^+e^-}[J/\psi(1S)]=$0.44, in good
agreement with the experimental value 0.41$\pm$0.07, what gives us 
confidence about the correct description of the $\eta_c(2S)$.

The most recently discovered charmonium state is the $X(3872)$, reported
by Belle \cite{chi1} in the $J/\Psi\pi^+\pi^-$ invariant mass distribution of the
$B^\pm\to K^\pm J/\psi\pi^+\pi^-$ reaction, and confirmed by the
CDF collaboration at Fermilab \cite{chi2}. Both experiments report a
similar mass, $3872.0\pm0.6\pm0.5$ MeV in Ref. \cite{chi1} and $3871.4\pm0.7\pm0.4$ 
MeV in Ref. \cite{chi2}, very close to the $D^0D^{*0}$ threshold ($3871.5\pm0.5$ MeV).

The proposed interpretations for the $X(3872)$ include $1^3D_3$, $1^3D_2$,
$1^1D_2$, $2^3P_1$ and $2^1P_1$ $c\bar c$ states. None of them
comfortably fit the observed properties of this state and therefore a
considerable experimental uncertainty still remains. The results of our model
for the possible quantum numbers are shown in Table \ref{t55}.
Although the well-established $1^3D_1$ is well reproduced by our model the predicted $1D$
candidates lie 70$-$80 MeV below its experimental mass whereas the $2P$ states
are 40 MeV above. These results clearly show that the
theoretical splitting in the $^3D_J$ multiplet is much smaller [$\sim$25 MeV] than the one necessary to correctly described
the $X(3872)$ and the $\Psi(3770)$ [$\sim$100 MeV].
Similar results have been found in other potential models which
suggests that the $X(3872)$ could present 
a more involved structure \cite{chi3,mes4}.

\subsection{Bottomonium and open beauty states}

A pretty good description of the experimental states shown in Tables \ref{t6}
and \ref{t7} is obtained. Some caution is necessary with respect of some
of the experimental data reported in Table \ref{t7}. The most important
discrepancy observed between the data and our results arise from
the $\eta_b(1S)$, all the other states being described with similar
accuracy to any other spectroscopic model designed to study a 
particular sector \cite{pedr}. However, the result for this state
is based in only one experimental work where
only one event has been observed \cite{pdgb},
and therefore this data needs confirmation.
The uncertainty on this data can be easily understood
from the surprisingly large spin splitting that it would
produce: $m_{\Upsilon} - m_{\eta_b} = 160$ MeV.
Such an energy difference clearly spoils the evolution of the
spin splittings with the mass of the quarks:
$m_\rho - m_\pi = 630$ MeV, $m_{K^*} - m_K = 397$ MeV,
$m_{D^*} - m_{D} = 146$ MeV, $m_{J/\Psi} - m_{\eta_c} =  117$ MeV,
$m_{B^*} - m_B = 46$ MeV, $m_{\Upsilon} - m_{\eta_b} = 160$ MeV.
The $S$ or $D$ assignment of $1^{--}$ high energy
$b\bar b$ excitations is not conclusive. Let us mention again 
that with the noncentral terms of our 
interaction a correct description of the hyperfine splittings
both in order and magnitude is obtained.
For the open-beauty mesons (Table \ref{t6}) we find theoretical states 
corresponding to $B^*_J(5732)$ and $B^*_{sJ}(5850)$, in both cases being 
$J^P=2^+$. 

\subsection{$J^{PC}=0^{++}$ states}
\label{scalar}

It is still not clear which are the members of the $0^{++}$ nonet
corresponding to the $L=S=1$ $n\bar n$ and $s\bar s$  multiplets.
There are too many 0$^{++}$ mesons observed in the
region below 2 GeV to be explained as $q\bar q$ states.
There have been reported in the PDG
two isovectors: $a_{0}(980)$ and $a_{0}(1450)$; five isoscalars:
$f_{0}(600), f_{0}(980), f_{0}(1370), f_{0}(1500)$ and $f_{0}(1710)$;
and three $I=1/2$: $K_{0}^*(1430)$, $K_{0}^*(1950)$ and recently
$\kappa (800)$. The quark model predicts the existence
of one isovector, two isoscalars and two $I=1/2$ states for each nonet.
Our results are shown in Table \ref{t8}.
Using this table one can try to assign the physical states
to $0^{++}$ nonet members. 

Let us discuss each state separately. With respect to the
isovector states, there is a candidate
for the $a_{0}(980)$, the $^3P_0$ member of the lowest $^3P_J$
isovector multiplet. The other candidate, the $a_{0}(1450)$,
is predicted to be the scalar member of the $2\,^3P_J$ excited isovector
multiplet. This reinforces the predictions of the quark model, the spin-orbit
force making lighter the $J=0$ states with respect to the $J=2$.
The assignment of the $a_0(1450)$
as the scalar member of the lowest $^3P_J$ multiplet \cite{ZZZZ}
would contradict this idea, because the $a_2(1320)$ is well
established as a $q \bar q$ pair. The same behavior is evident in the
$c\bar c$ and $b\bar b$ spectra, making impossible to
describe the $a_0(1450)$ as a member of the lowest $^3P_J$
isovector multiplet without spoiling the description of heavy-quark multiplets.
However, in spite of the correct description of the mass
of the $a_{0}(980)$, the model predicts a pure light-quark content,
what seems to contradict some experimental observations \cite{XXXX}.
The $a_{0}(1450)$ is predicted to be also a pure light-quark structure
obtaining a mass somewhat higher than the experiment.
		   
In the case of the isoscalar states, we find a candidate for the
$f_0(600)$ with a strangeness content around
$10\%$. There are no $I=0$ states with a mass close to 1 or 1.5 GeV,
which would correspond to the $f_{0}(980)$ and the $f_{0}(1500)$,
and they cannot be found for any
combination of the model parameters. It seems that a different 
structure rather than a naive $q\bar q$ pair is needed.
In particular, the $f_0(1500)$ is a clear
candidate for the lightest glueball \cite{amsl}, while the $f_0(980)$
has been suggested as a possible four-quark state \cite{tera,vij2} 
what would make it compatible with the similar branching ratios observed 
for the $J/\psi \to f_0(980)\phi $ and $J/\psi \to f_0(980)\omega$ 
decays \cite{XXXX}. Concerning the $f_{0}(1370)$ (which may actually
correspond to two different states \cite{blac}) we obtain two 
states around this energy, the heavier one with a dominant nonstrange
content which favors its assignment to the $f_0(1370)$;  
the other with a high $s\bar s$ content without having an experimental
partner. Let us however remember that in this energy region there is a state,
the $f_2(1430)$, that does not fit into the isoscalar $2^{++}$ sector 
and has been recently suggested as a possible $0^{++}$ state \cite{vlad}.
Finally a dominant $n\bar n$ state corresponding to the $f_0(1710)$ is 
obtained. Our results concluding that $f_0(1370)$, $f_0(1500)$ and 
$f_0(1710)$ are dominantly $n\bar n$, non $q \bar q$, and $n\bar n$ 
respectively differ from the conclusion of 
Refs. \cite{amsl,AMS2} obtaining that 
$f_0(1710)$ is dominantly $s \bar s$ and are also in contrast 
to the predictions of Ref. \cite{WEIN} which prefers to 
assign $f_0(1500)$ to an $s\bar s$ state and $f_0(1710)$ to a 
glueball. This makes clear the complicated situation in the scalar sector
with several alternative interpretations of the observed states.
The study of radiative transitions and two photon decay widths 
should help to understand the flavor mixing and the nature of the $I=0$ scalar
sector. We obtain two states around 1.9 GeV, 
a $2 \, ^3P_0$ state with a dominant $s \bar s$ content 
and a $4 \, ^3P_0$ with a dominant $n \bar n$ content.
Being the $f_0(2020)$ an experimentally known
$n \bar n$ meson, its identification with the
$4 \, ^3P_0$ state is clear. Therefore, we find an unobserved
$s \bar s$ scalar with a mass around 1.9 GeV as has also been
suggested in Ref.  \cite{brna}. Finally, although for consistency
not given in the tables, we find a candidate
for the $3 \, ^3P_0$ state $f_0(2200)$, experimentally identified as an 
$s \bar s$ state \cite{ani2}, with an energy of 2212 MeV.

Concerning the $I=1/2$ sector, as a consequence of the larger mass
of the strange quark as compared to the light ones, our model always
predicts a mass for the lowest $0^{++}$ state 200 MeV greater than the
$a_{0}(980)$ mass. Therefore, being the $a_0(980)$ the
member of the lowest isovector scalar multiplet, the $\kappa (800)$
cannot be explained  as a $q\bar q$ pair.  
We find a candidate for the $K_0^*(1430)$ although with a smaller mass.  

Our results indicate that the light-scalar
sector cannot be described in a pure $q\bar q$ scheme and more
complicated structures or mixing with multiquark system seems to be
needed. Concerning the $f_0(980)$ and the $\kappa(800)$ our conclusions
are very similar to those obtained in Ref. \cite{umek} using the extended
Nambu-Jona-Lasinio model in an improved ladder approximation of the 
Bethe-Salpeter equation. This seems to indicate that relativistic
corrections would not improve the situation and the conclusions remain model 
independent.     

One finds in the literature several alternative explanations to understand
the rather complicated scenario of the scalar mesons. An earlier
attempt to link the understanding of the $NN$ interactions with meson
spectroscopy was done based on the J\"ulich potential model \cite{PRD52}. The
structure of the scalar mesons $a_0(980)$ and $f_0(980)$ was investigated in
the framework of a meson exchange model for $\pi\pi$ and $\pi\eta$ scattering.
The $K\bar K$ interaction generated by the vector-meson exchange, which for
isospin $I=0$ is strong enough to generate a bound state is much weaker for
$I=1$, making a degeneracy of $a_0(980)$ and $f_0(980)$ impossible, as found in
our model. Although both scalar mesons
result from the coupling to the $K\bar K$ channel explaining in a natural way
their similar properties, the underlying
structure obtained was, however, quite different. Whereas the $f_0(980)$
appears to be a $K\bar K$ bound state the
$a_0(980)$ was found to be a dynamically generated threshold effect.

In a different fashion within the quark model the same problem was illustrated
in Ref.\cite{TORN}. The bare mass used for the $n\bar n$ pair is much larger
than the $a_0(980)$ and $f_0(980)$ experimental masses. It is the effect of the
two-pseudoscalar meson thresholds the responsible for the substantial shift to a
lower mass than what is naively expected from the $q\bar q$ component alone.
This gives rise to an important $K\bar K$ and $\pi\eta'$ components in the
$a_0(980)$ and $K\bar K$,  $\eta\eta$, $\eta'\eta'$ and $\eta\eta'$ in the
$f_0(980)$. In particular for the $a_0(980)$ they obtain the $K\bar K$
component to be dominant near the peak, being about $4-5$ times larger than the
$q\bar q$ component. A similar conclusion, that the description of the
$a_0(980)$ and $f_0(980)$ requires from more complex structures, is also obtained from our
analysis. The absence of the $f_0(1500)$ in our $q\bar q$ scheme
make also contact with the indication that this state could correspond to the
lightest scalar glueball \cite{PENI}.

A similar problem as the one observed with the light scalars appeared in the
open charm sector. As we have already discussed, the two recently measured 
$0^+$ states do not fit into a $q\bar q$ description. 
Possible alternatives to understand
their masses, as for example being $D K$ molecules \cite{bacl} or
tetraquarks \cite{tera}, have been suggested. Using the
same interacting hamiltonian presented in this work an estimation of the 
lowest scalar open charm tetraquark has been done in Ref. \cite{vij2}. 
The variational estimation was performed under the assumption that
internal orbital angular momentum does not give an important contribution,
what has been strictly tested in the case of heavy-light tetraquarks 
\cite{vij4}, but it is well known to influence
light-quark systems. Having in mind this precaution a mass of 2389 MeV was 
obtained for the light open-charm $[(ns)(\bar n \bar c)]$ 
tetraquark that could very well correspond to the $D^*_{sJ}(2317)$. 
A consistent calculation of four light-quark structures seems to be 
advocated for a full understanding of the scalar sector.

\section{Summary}

We have performed an exhaustive study of the meson spectra from the 
light $n \bar n$ states to the $b \bar b$ mesons within the same model.
The quark-quark interaction takes into account QCD perturbative
effects by means of the one-gluon-exchange potential and the most important 
nonperturbative effects through the hypothesis of a screened confinement 
and the spontaneous breaking of chiral symmetry.
The model incorporates in a natural way the
$n \bar n \leftrightarrow s \bar s$ isoscalar mixing due to the $V_K$ 
potential, and the $1^+$ $S=0$ and $S=1$ mixing 
caused by the spin-orbit force. 
An arbitrary rate of scalar/vector confinement has been used, finding 
evidence of a strong scalar component. Annihilation effects have not been 
taken into account although their contribution seems to be important for
the description of the $1^{+-}$ light isoscalar ground state.
We have obtained a reasonable description of most part of the
well established $q \bar q$ states for all flavors. The success
of the model allowed us to make predictions with respect to those 
states whose quantum numbers, existence or nature is under debate.

In the light-isoscalar sector our results support the speculation 
pointed out by the PDG about the possible nonexistence of the
$f_1(1420)$. We do not find a theoretical $2^{++}$ state 
corresponding to the $f_2(1430)$, in agreement with recent 
experiments that opened the possibility of a different assignment
of quantum numbers for this meson being a $0^{++}$ state. 
Our model predicts a scalar meson in this energy region without
an experimental partner.
Besides, we have only found one $q \bar q$ state in
the 1.6 GeV region, which seems to indicate that either the
$f_2(1565)$ and the $f_2(1640)$ are the same state 
as suggested by the PDG, or that the $f_2(1565)$ goes beyond 
the naive $q\bar q$ structure. In the light-scalar sector
it seems very difficult to accommodate the $f_0(980)$, 
the $\kappa(800)$ and the $f_0(1500)$ in a $q \bar q$ scheme. 
In the light-strange sector we do not find a $q \bar q$
state to be identified with the $K^*(1410)$, favoring its
possible hybrid structure. The same situation occurs for the
$K_2(1580)$, although in this case the poor experimental data 
do not assure its existence. 

Concerning the flavor content of the $\eta(1440)$ our model indicates 
that it is a dominant $n\bar n$ state, with a probability of 70.3 \%.
The $\phi_3(1875)$ is compatible with an $s\bar s$ content but there 
should exist an unobserved $n\bar n$ state in the same energy region. 
We have also found evidence of the existence of a $1^{+-}$
light-isoscalar meson with a dominant $s \bar s$ content
and a mass around 1.97 GeV, and an isoscalar $s \bar s$ $2^{+-}$
state at 1.85 GeV. Finally, our model assigns
the quantum numbers $1^-$ to $K(1630)$.

For the heavy-quark sector the experimental situation is changing very
fast. New experiments and reanalysis of old data are being done and new
states being discovered. Some of them fit nicely in a $q\bar q$ scheme,
but others are impossible to accommodate. In the open charm
sector, the $D^*(2640)$ and $D_{sJ}(2573)$ are compatible with $c \bar n$
and $c \bar s$ mesons with quantum numbers 
$0^-$ and $2^+$, respectively. However,
the recently discovered $0^+$ states, the $D^*_{sJ}(2317)$ and
$D_J(2308)$, seem to have a completely different structure. We have argued 
a possible explanation to describe their low masses based in
a tetraquark structure that may
have positive parity without orbital excitation. 
Finally, there is an obvious identification of a $q \bar q$ state
for the $h_c(1P)$ with quantum numbers $1^{+-}$.

We consider that this type of study based on models whose parameters are
constrained in the description of other low-energy systems should be a 
complementary useful tool to deepen the understanding of the meson spectra.
The next step in this effort to a comprehensive description of the new
data concerning the meson spectra should be the analysis of the electroweak
and strong decays of mesons that will be the subject of a future 
publication \cite{futu}.

\section{ acknowledgments}

This work has been partially funded by Ministerio 
de Ciencia y Tecnolog{\'{\i}}a under Contract No. BFM2001-3563, 
by Junta de Castilla y Le\'{o}n under Contract No. SA-104/04.

\begin{table}[tbp]
\caption{Quark model parameters.}
\label{t1}
\begin{center}
\begin{tabular}{cc|ccc}
&&$m_u=m_d$ (MeV) & 313 & \\ 
&Quark masses&$m_s$ (MeV)     & 555 & \\ 
&&$m_c$ (MeV)     & 1752& \\
&&$m_b$ (MeV)     & 5100& \\ 
\hline
&&$m_{\pi}$ (fm$^{-1}$) & 0.70&\\
&&$m_{\sigma}$ (fm$^{-1}$)& 3.42&\\ 
&&$m_{\eta}$ (fm$^{-1}$)  & 2.77&\\ 
&Goldstone bosons&$m_K$ (fm$^{-1}$)       & 2.51&\\ 
&&$\Lambda_{\pi}=\Lambda_{\sigma}$ (fm$^{-1}$) & 4.20 &\\
&&$\Lambda_{\eta}=\Lambda_K$ (fm$^{-1}$) & 5.20&\\ 
&& $g_{ch}^2/(4\pi)$      & 0.54&\\ 
&&$\theta_P(^o)$          & $-$15&\\ 
\hline
&&$a_c$ (MeV)             &430&\\
&Confinement&$\mu_c$ (fm$^{-1}$)&0.70&\\ 
&&$\Delta$ (MeV)          &181.10&\\
&&$a_s$                   &0.777&\\ 
\hline
&&$\alpha_0$              &2.118&\\
&&$\Lambda_0$ (fm$^{-1}$) &0.113&\\
&OGE&$\mu_0$ (MeV)        &36.976&\\
&&$\hat r_0$ (MeV fm)     &28.170&\\
&&$\hat r_g$ (MeV fm)     &34.500&\\
\end{tabular}
\end{center}
\end{table}

\begin{table}[tbp]
\caption{Energies (in MeV) obtained when solving the
Schr\"odinger equation for two different partial waves when
the different noncentral terms are switched on.}
\label{tnw}
\begin{center}
\begin{tabular}{lcc}
Potential & $^3D_1$ (I=1) & $^3P_0 (n\overline n)$ (I=0) \\
\hline
$V^C_{qq} (\vec{r}_{ij}) + V^C_{OGE} (\vec{r}_{ij}) + V^C_{CON} (\vec{r}_{ij})$
&   1602  & 1261 \\
$+ V^T_{qq} (\vec{r}_{ij})$                   &   1598  & 1008 \\
$ + V^{SO}_{OGE} (\vec{r}_{ij})$               &   1474  &  225 \\
$ + V^{SO}_{CON} (\vec{r}_{ij})$               &   1508  &  335 \\
$ + V^{SO}_{\sigma} (\vec{r}_{ij})$     
&   1518  &  500 \\
\end{tabular}
\end{center}
\end{table}
\begin{table}[tbp]
\caption{Masses, in MeV, of $I=0$ light-quark mesons up to 2 GeV. QM denotes
the results of the present model and Flavor stands for the
dominant component of the flavor wave function. Experimental data, PDG, 
are taken from Ref. \protect\cite{pdgb}. In the second column we 
denote by a question mark 
those states whose existence is not clear. In the third column, if there are
several candidates for an experimental state, we underline our preferred
assignment. See text for details.}
\label{t2}
\begin{center}
\begin{tabular}{ccccc}
 $(nL) \, J^{PC}$ & State & QM & Flavor & PDG \\ 
\hline
$(1S) \, 0^{-+}$ & $\eta$ & 572& $(n\bar n)$ & 547.75$\pm0.12$ \\ 
$(1S) \, 0^{-+}$ & $\eta^{\prime}(958)$ & 956 &$(s\bar s)$& 957.8$\pm0.1$ \\ 
$(2S) \, 0^{-+}$ & $\eta(1295)$ & 1290 &$(n\bar n)$ & 1294$\pm4$ \\ 
$(2S) \, 0^{-+}$ & $\eta(1760)$ & 1795 &$(s\bar s)$ & 1760$\pm$11 \\ 
$(3S) \, 0^{-+}$ & $\left[\begin{array}{c} \eta(1405)\\ \eta(1475)\end{array}
\right]$ & 1563 &  $(n\bar n)$&$\left[\begin{array}{c}  1410.3\pm2.6\\1476\pm4
\end{array}\right]$ \\ 
$(1D) \, 2^{-+}$ & $\eta_2(1645)$ & 1600 &$(n\bar n)$ & 1617$\pm$5 \\ 
$\left[\begin{array}{c} (1D) \, 2^{-+}\\(2D)\,2^{-+}\end{array}\right]$ 
& $\eta_2(1870)$ &$\left[\begin{array}{c} 1853\\ {\underline{1863}}
\end{array}\right]$ & $\left[\begin{array}{c} (s\bar s) \\ {\underline{
(n\bar n)}} \end{array}\right]$ & 1842$\pm$8 \\ 
$(1S) \, 1^{--}$ & $\omega(782)$ & 691 &$(n\bar n)$ & 782.54$\pm0.11$ \\ 
$(1S) \, 1^{--}$ & $\phi(1020)$ & 1020 &$(s\bar s)$ & 1019.46$\pm0.02$ \\ 
$(2S) \, 1^{--}$ & $\omega(1420)$ & 1444 &$(n\bar n)$ & 1400$-$1450\\ 
$(2S) \, 1^{--}$ & $\phi(1680)$ & 1726 &$(s\bar s)$ & 1680$\pm$20 \\ 
$(3S) \, 1^{--}$ & $\omega(1650)$ & 1784 &$(n\bar n)$ & 1670$\pm$30 \\ 
$(1D) \, 3^{--}$ & $\omega_3(1670)$ & 1631 &$(n\bar n)$ & 1667$\pm4$ \\ 
$\left[\begin{array}{c} (1D) \, 3^{--}\\ (2D) \, 3^{--}\end{array}\right]$ 
& $\phi_3(1850)$ & $\left[\begin{array}{c} {\underline{1875}}\\ 1876
\end{array}\right]$ &
$\left[\begin{array}{c} {\underline{(s\bar s)}}\\(n\bar n)
\end{array}\right]$ & 1854$\pm7$ \\ 
$(1P) \, 1^{+-}$ & $h_1(1170)$ & 1257 &$(n\bar n)$ & 1170$\pm$20 \\ 
$(1P) \, 1^{+-}$ & $h_1(1380)$ & 1511 &$(s\bar s)$ & 1386$\pm$19 \\ 
$(2P) \, 1^{+-}$ & $h_1(1595)$ & 1700 &$(n\bar n)$ & $1594^{+18}_{-62}$ \\ 
$(2P) \, 1^{+-}$ &  & 1973 &$(s\bar s)$ & \\ 
$(1P) \, 2^{++}$ & $f_2(1270)$ & 1311 &$(n\bar n)$ & 1275.4$\pm1.2$ \\ 
$(1P) \, 2^{++}$ & $f_2^{\prime}(1525)$ & 1556 &$(s\bar s)$ & 1525$\pm$5 \\ 
$(-) \, 2^{++}$ & $f_2(1430)$ & $-$ & $-$ & $\approx$1430 \\ 
$(2P) \, 2^{++}$ & $\left[\begin{array}{c} f_2(1565)?\\f_2(1640)\end{array}
\right]$ & 1725 &  $(n\bar n)$&$\left[\begin{array}{c}  1546\pm12\\1638\pm6
\end{array}\right]$ \\ 
$(1F) \, 2^{++}$ & $f_2(1810)$ & 1789 &$(n\bar n)$ & 1815$\pm$12 \\ 
$(3P) \, 2^{++}$ & $f_2(1910)$ & 1906 &$(n\bar n)$ & 1915$\pm$7 \\ 
$(2P) \, 2^{++}$ & $f_2(1950)$ & 1999 &$(s\bar s)$& 1934$\pm$12 \\ 
$(1P) \, 1^{++}$ & $f_1(1285)$ & 1271 &$(n\bar n)$ & 1281.8$\pm0.6$ \\ 
$(1P) \, 1^{++}$ & $\left[\begin{array}{c} f_1(1420)?\\f_1(1510)\end{array}
\right]$ & 1508& $(s\bar s)$ & $\left[\begin{array}{c} 1426.3\pm1.1\\
1518\pm5 \end{array}\right]$\\ 
\end{tabular}
\end{center}
\end{table}

\begin{table}[tbp]
\caption{Masses, in MeV, of $I=1$ light-quark mesons up to 2 GeV. 
QM denotes the results of the present model. 
Experimental data (PDG) are taken from Ref. \protect\cite{pdgb}.
We denote by a dagger those PDG states whose masses are not explicitly given. 
We use for them the more recent experimental data.}
\label{t2b}
\begin{center}
\begin{tabular}{cccc}
 $(nL) \, J^{PC}$ & State & QM & PDG \\ 
\hline
$(1S) \, 0^{-+}$ & $\pi$ & 139 & 139$$ \\ 
$(2S) \, 0^{-+}$ & $\pi(1300)$ & 1288 & 1300$\pm$100 \\ 
$(3S) \, 0^{-+}$ & $\pi(1800)$ & 1720 & 1812$\pm$14 \\ 
$(1D) \, 2^{-+}$ & $\pi_2(1670)$ & 1600 & 1672.4$\pm$3.2 \\ 
$(1S) \, 1^{--}$ & $\rho(770)$ & 772 & 775.8$\pm0.5$ \\ 
$(2S) \, 1^{--}$ & $\rho(1450)$ & 1478 & 1465$\pm$25 \\ 
$\left[\begin{array}{c} (3S) \, 1^{--}\\(2D)\,1^{--}\end{array}\right]$ 
& $\rho(1700)$ &$\left[\begin{array}{c} 1802\\ 1826
\end{array}\right]$ & 1720$\pm$20 \\ 
$(4S) \, 1^{--}$ & $\rho(1900)$ & 1927 & 1911$\pm$5$^\dag$ \\ 
$(1D) \, 3^{--}$ & $\rho_3(1690)$ & 1636 & 1691$\pm5$ \\ 
$(2D) \, 3^{--}$ & $\rho_3(1990)$ & 1878 & $1981\pm14^\dag$ \\ 
$(1P) \, 1^{+-}$ & $b_1(1235)$ & 1234 & 1229.5$\pm3.2$ \\ 
$(1P) \, 1^{++}$ & $a_1(1260)$ & 1205 & 1230$\pm$40 \\ 
$(2P) \, 1^{++}$ & $a_1(1640)$ & 1677 & 1647$\pm$22 \\ 
$(1P) \, 2^{++}$ & $a_2(1320)$ & 1327 & 1318.3$\pm0.6$ \\ 
$(2P) \, 2^{++}$ & $a_2(1700)$ & 1732 & 1732$\pm16$ \\ 
\end{tabular}
\end{center}
\end{table}

\begin{table}[tbp]
\caption{Masses, in MeV, of the light-strange mesons up to 2 GeV. QM denotes
the results of the present model. Experimental data (PDG) are taken 
from Ref. \protect\cite{pdgb}. We denote by a dagger those PDG
states whose masses are not explicitly given. We take 
for them the more recent experimental data.
The spin is indicated because $C$ parity is not well defined. 
In those cases where the PDG does not give the $J^P$ quantum numbers
and we find a candidate for the state, we quote between square brackets
our predictions. QM(mixed) denotes the mass of the 
states after being mixed according to Eq. (\protect\ref{mix}).}
\label{t3}
\begin{center}
\begin{tabular}{cccccc}
$(nL) \, J^{P}$ & Spin & State & QM & QM(mixed) & PDG \\ 
\hline
$(1S) \, 0^{-}$ & 0 & $K$ & 496 && 495 \\ 
$(2S) \, 0^{-}$ & 0 & $K(1460)$ & 1472 && $\approx1460^\dag$ \\ 
$(3S) \, 0^{-}$ & 0 & $K(1830)$ & 1899 && $\approx1830^\dag$ \\ 
$(1S) \, 1^{-}$ & 1 & $K^*(892)$ & 910 && 891.7$\pm0.3$ \\ 
$(-)  \, 1^{-}$ &$-$& $K^*(1410)$ &$-$&& 1414$\pm$15 \\ 
$(1D) \, 1^{-}$ & 1 & $K^*(1680)$ & 1698 && 1717$\pm27$ \\ 
$(2S) \, ?^{?}\,[1^-]$ &1& $K(1630)$ &1620&& 1629$\pm$7 \\ 
$(1P) \, 1^{+}$ & 1 & $K_1(1270)$ & 1372 &1352& 1273$\pm7$ \\ 
$(1P) \, 1^{+}$ & 0 & $K_1(1400)$ & 1394 &1414& 1402$\pm7$ \\ 
$(2P) \, 1^{+}$ & 1 & $K_1(1650)$ & 1841 &1836& 1650$\pm50$ \\ 
$(2P) \, 1^{+}$ & 0 &  & 1850 & 1856 &  \\ 
$(1P) \, 2^{+}$ & 1 & $K^*_2(1430)$ & 1450 && 1425.6$\pm1.5$ \\ 
$(1F) \, 2^{+}$ & 1 & $K^*_2(1980)$ & 1968 && 1973$\pm$26 \\ 
$(-)  \, 2^{-}$ &$-$& $K_2(1580)$ & $-$ && $\approx1580^\dag$ \\ 
$(1D) \, 2^{-}$ & 1 & $K_2(1770)$ & 1741 & 1709 & 1773$\pm$8 \\ 
$(1D) \, 2^{-}$ & 0 & $K_2(1820)$ & 1747 & 1779 & 1816$\pm$13 \\ 
$(1D) \, 3^{-}$ & 1 & $K^*_3(1780)$ & 1766 && 1776$\pm7$ \\ 
\end{tabular}
\end{center}
\end{table}

\begin{table}[tbp]
\caption{Same as Table \protect\ref{t3} for $D$ and $D_s$ mesons.}
\label{t4}
\begin{center}
\begin{tabular}{c|cccccc}
Meson &$(nL) \, J^{P}$ & Spin & State & QM & QM(mixed)& PDG \\ 
\hline
&$(1S) \, 0^{-}$ & 0 & $D$ & 1883 && 1867.7$\pm0.5$ \\ 
&$(1S) \, 1^{-}$ & 1 & $D^*$ & 2010 && 2008.9$\pm$0.5 \\ 
$D$&$(1P) \, 1^{+}$ & 0 & $D_1(2420)$ & 2492 &2454& 2425$\pm$4 \\ 
&$(1P) \, 2^{+}$ & 1 & $D_2^*(2460)$ & 2502 && 2459$\pm$4 \\ 
&$(2S) \, ?^?\,[0^{-}]$ & 0 & $D^*(2640)$ & 2642& & 2637$\pm$7 \\ 
\hline
&$(1S) \, 0^{-}$ & 0 & $D_s$ & 1981 && 1968.5$\pm$0.6\\ 
&$(1S) \, 1^{-}$ & 1 & $D_s^*$ & 2112 && 2112.4$\pm$0.7 \\ 
$D_s$&$(1P) \, 0^{+}$ & 1 & $D^*_{sJ}(2317)$ & 2469 && 2317.4$\pm$0.9\\ 
&$(1P) \, 1^{+}$ & 0 & $D^*_{sJ}(2460)$ & 2550 & 2543 & 2459.3$\pm$1.3\\ 
&$(1P) \, 1^{+}$ & 1 & $D_{s1}(2536)$ & 2563 &2571& 2535.3$\pm$0.6 \\ 
&$(1P) \, ?^?\,[2^{+}]$ & 1 & $D_{sJ}(2573)$ & 2585 && 2572.4$\pm$1.5 \\ 
\end{tabular}
\end{center}
\end{table}

\begin{table}[tbp]
\caption{Masses, in MeV, of the first positive parity $D$ and $D_s$ mesons 
compared to recently measured experimental data. QM denotes the results
of the present model. QM(mixed) indicates the mass of the states
after being mixed according to Eq. (\protect\ref{mix}).}
\label{t9}
\begin{center}
\begin{tabular}{c|ccccc}
Meson &$(nL)$ $J^P$&$(1P)$ $0^+$ &$(1P)$ $1^+$ &$(1P)$ $1^+$ &$(1P)$ $2^+$ \\ 
\hline
&PDG\protect\cite{pdgb}&&& 2425$\pm$4 & 2459$\pm$4  \\
&FOCUS\protect\cite{focu,foc2}&$\sim 2420$ &  &  & 2468$\pm$2 \\
&Belle\protect\cite{bel5}&2308$\pm$36 & 2427$\pm$36 &2421$\pm$2&2461$\pm$4\\
$D$&CLEO\protect\cite{cleo4}&  & 2461$\pm$51 &  &   \\
&DELPHI\protect\cite{delp}&  & 2470$\pm$58 &  &   \\
&&  &  &  &    \\
&QM& 2436 & 2496 & 2492 & 2502  \\
&QM(mixed)&  &2535&2454& \\
\hline
&PDG\protect\cite{pdgb}& 2317.4$\pm$0.9  & 2459.3$\pm$1.3 & 2535.3$\pm$0.6 & 2572.4$\pm$1.5  \\
&FOCUS\protect\cite{focu,foc2}&  &  & 2535.1$\pm$0.3 & 2567.3$\pm$1.4  \\
&BaBar\protect\cite{baba,bab2}& 2317.3$\pm$0.9 
	&2458.0$\pm$1.4&  &   \\
$D_s$&CLEO\protect\cite{cleo,cleo2,cleo3} & 2318.1$\pm$1.2 
	&2463.1$\pm$2.0&  &   \\
&Belle\protect\cite{bel3,bel4}& 2319.8$\pm$2.1 
	&2456.5$\pm$1.8&  &   \\
&&  &  &  &   \\
&QM       & 2469 &2563  &2550 & 2585  \\
&QM(mixed)&      &2571  &2543  & \\
\end{tabular}
\end{center}
\end{table} 

\begin{table}[tbp]
\caption{Same as Table \protect\ref{t2} for charmonium.}
\label{t5}
\begin{center}
\begin{tabular}{cccc}
$(nL) \, J^{PC}$ & State & QM & PDG \\ 
\hline
$(1S) \, 0^{-+}$ & $\eta_c(1S)$ & 2990 & 2979.6$\pm1.2$ \\ 
$(1S) \, 1^{--}$ & $J/\psi(1S)$ & 3097 & 3096.916$\pm0.011$ \\ 
$(1P) \, 0^{++}$ & $\chi_{c0}(1P)$ & 3436 & 3415.19$\pm0.34$ \\ 
$(1P) \, 1^{++}$ & $\chi_{c1}(1P)$ & 3494 & 3510.59$\pm0.10$ \\ 
$(1P) \, 2^{++}$ & $\chi_{c2}(1P)$ & 3526 & 3556.26$\pm0.11$ \\ 
$(1P) \, ?^{??}\,[1^{+-}]$ & $h_c(1P)$ & 3507 & 3526.21$\pm$0.25 \\ 
$(2S) \, 0^{-+}$ & $\eta_c(2S)$ & 3627 & 3654$\pm$10 \\ 
$(2S) \, 1^{--}$ & $\psi(2S)$ & 3685 & 3686.093$\pm0.034$ \\ 
$(1D) \, 1^{--}$ & $\psi(3770)$ & 3775 & 3770.0$\pm2.4$ \\ 
$(1D) \, 2^{--}$ & $\psi(3836)$ & 3790 & 3836$\pm$13 \\ 
$(3S) \, 1^{--}$ & $\psi(4040)$ & 4050 & 4040$\pm$10 \\ 
$(2D) \, 1^{--}$ & $\psi(4160)$ & 4103 & 4159$\pm$20 \\ 
$\left[\begin{array}{c} (4S) \, 1^{--}\\ (3D) \, 1^{--}\end{array}\right]$
& $\psi(4415)$ & $\left[\begin{array}{c}  4307 \\ 4341\end{array}\right]$
 & 4415$\pm$6 \\ 
\end{tabular}
\end{center}
\end{table}

\begin{table}[tbp]
\caption{Masses predicted by our model for the $q\overline{q}$ states compatible with
the $X(3872)$}
\label{t55}
\begin{center}
\begin{tabular}{ccccc}
$1^3D_3$&$1^3D_2$&$1^1D_2$&$2^3P_1$&$2^1P_1$\\
\hline
3802&3790&3793&3913&3924\\
\end{tabular}
\end{center}
\end{table}

\begin{table}[tbp]
\caption{Same as Table \protect\ref{t3} for $B$, $B_s$ and $B_c$ mesons.}
\label{t6}
\begin{center}
\begin{tabular}{c|ccccc}
&$(nL) \, J^P$ & Spin & State & QM & PDG \\ 
\hline
&$(1S) \, 0^{-}$ & 0 & $B$ & 5281 & 5279.2$\pm$0.5 \\ 
$B$&$(1S) \, 1^{-}$ & 1 &$B^*$ & 5321 & 5325.0$\pm$0.6 \\ 
&$(1P) \, ?^?\,[2^{+}]$ & 1 & $B_J^*(5732)$ & 5790 & 5698$\pm$8 \\ 
\hline
&$(1S) \, 0^{-}$ & 0 & $B_s$ & 5355 & 5369.6$\pm$2.4 \\ 
$B_s$&$(1S) \, 1^{-}$ & 1 & $B_s^*$ & 5400 & 5416.6$\pm$3.5 \\ 
&$(1P) \, ?^?\,[2^{+}]$ & 1 & $B_{sJ}^*(5850)$ & 5855 & 5853$\pm$15\\ 
\hline
$B_c$&$(1S) \, 0^{-}$ & 0 & $B_c$ & 6277 & 6400$\pm$410 \\ 
\end{tabular}
\end{center}
\end{table}

\begin{table}[tbp]
\caption{Same as Table \protect\ref{t2} for bottomonium. We denote by 
an asterisk a experimental state recently reported in Ref.\protect\cite{clen}.}
\label{t7}
\begin{center}
\begin{tabular}{cccc}
$(nL) \, J^{PC}$ & State & QM & PDG \\ 
\hline
$(1S) \, 0^{-+}$ & $\eta_b(1S)$ & 9454 & 9300$\pm$28 \\ 
$(1S) \, 1^{--}$ & $\Upsilon(1S)$ & 9505 & 9460.30$\pm0.26$ \\ 
$(1P) \, 0^{++}$ & $\chi_{b0}(1P)$ & 9855 & 9859.9$\pm$1.0 \\ 
$(1P) \, 1^{++}$ & $\chi_{b1}(1P)$ & 9875 & 9892.7$\pm$0.6 \\ 
$(1P) \, 2^{++}$ & $\chi_{b2}(1P)$ & 9887 & 9912.6$\pm$0.5 \\ 
$(2S) \, 1^{--}$ & $\Upsilon(2S)$ & 10013 & 10023.26$\pm0.31$ \\ 
$(1D) \, 2^{--}$ & $\Upsilon(1D_2)^*$ & 10119 & 10162.2$\pm1.6^*$\\ 
$(2P) \, 0^{++}$ & $\chi_{b0}(2P)$ & 10212 & 10232.1$\pm$0.6 \\ 
$(2P) \, 1^{++}$ & $\chi_{b1}(2P)$ & 10227 & 10255.2$\pm$0.5 \\ 
$(2P) \, 2^{++}$ & $\chi_{b2}(2P)$ & 10237 & 10268.5$\pm$0.4 \\ 
$(3S) \, 1^{--}$ & $\Upsilon(3S)$ & 10335 & 10355.2$\pm3.5$ \\ 
$(4S) \, 1^{--}$ & $\Upsilon(4S)$ & 10577 & 10580.0$\pm3.5$ \\ 
$\left[\begin{array}{c} (5S) \, 1^{--}\\(4D) \, 1^{--}\end{array}\right]$ 
& $\Upsilon(10860)$ &$\left[\begin{array}{c} 10770\\10803\end{array}\right]$
& 10865$\pm$8 \\ 
$\left[\begin{array}{c} (6S) \, 1^{--}\\(5D) \, 1^{--}\end{array}\right]$ 
& $\Upsilon(11020)$ &$\left[\begin{array}{c} 10927\\10953\end{array}\right]$
& 11019$\pm$8 \\ 
\end{tabular}
\end{center}
\end{table}

\begin{table}[tbp]
\caption{Same as Table \protect\ref{t2} for the light-scalar mesons. We
have included the $\kappa(800)$, whose isospin is not well determined
being preferred $I=1/2$ \protect\cite{kapa}. }
\label{t8}
\begin{center}
\begin{tabular}{c|ccccc}
&$(nL) \, J^{PC}$ & State & QM  & Flavor & PDG \\ 
\hline
$I=1$&$(1P) \, 0^{++}$ & $a_0(980)$ & 984 &$(n\bar n)$& 984.7$\pm$1.2 \\ 
&$(2P) \, 0^{++}$ & $a_0(1450)$ & 1587 &$(n\bar n)$& 1474$\pm$19 \\ 
\hline
&$(1P) \, 0^{++}$ & $f_0(600)$ & 413 &$(n\bar n)$& 400$-$1200 \\ 
&$(-) \, 0^{++}$ & $f_0(980)$ &  $-$ &$-$& 980$\pm$10 \\ 
&$(2P) \, 0^{++}$ & $f_0(1370)$ & 1395 &$(n\bar n)$& 1200$-$1500 \\ 
$I=0$&$(1P) \, [0^{++}]$ & $-$  & 1340 &$(s\bar s)$& $-$ \\ 
&$(-) \, 0^{++}$ & $f_0(1500)$ &  $-$ &$-$& 1507$\pm$5 \\ 
&$(3P) \, 0^{++}$ & $f_0(1710)$ & 1754 &$(n\bar n)$& 1714$\pm$5 \\ 
&$\left[\begin{array}{c} (4P) \, 0^{++}\\ (2P) \, 0^{++}\end{array}\right]$ 
& $f_0(2020)$ & $\left[\begin{array}{c} {\underline{1880}}\\ 1894
\end{array}\right]$ &
$\left[\begin{array}{c} {\underline{(n\bar n)}}\\(s\bar s)
\end{array}\right]$ & 1992$\pm16$ \\ 
\hline
&$(-) \, 0^{+}$  & $\kappa(800)$ & $-$ &$-$& $\approx 800$ \\ 
$I=1/2$&$(1P) \, 0^{+}$  & $K^*_0(1430)$ & 1213 & $(n \bar s)$ & 1412$\pm$6 \\ 
&$(2P) \, 0^{+}$  & $K^*_0(1950)$ & 1768 &$(n\bar s)$& 1945$\pm$22 \\ 
\hline
\end{tabular}
\end{center}
\end{table}

\begin{table}
\begin{center}
\caption{Masses, in MeV, of the first positive parity $D$ and $D_s$ mesons 
with $a_s=0.46$. They are already mixed according to Eq. (\protect\ref{mix}).}
\label{t10}
\begin{tabular}{cc|cccc}
&Meson&$(nL)$ $J^P$&State&QM(mixed)&\\
\hline
&&$(1P)$ $0^+$&$D^*_{sJ}(2317)$ & 2317&\\
&$D_s$&$(1P)$ $1^+$& $D_{sJ}(2463)$ & 2482&\\
&&$(1P)$ $1^+$&$D_{s1}(2536)$ & 2574&\\
&&$(1P)$ $2^+$&$D_{sJ}(2573)$ & 2633&\\
\hline
&&$(1P)$ $0^+$&2308$\pm$36 & 2134&\\
&$D$&$(1P)$ $1^+$&$D_{1}(2420)$ & 2354&\\
&&$(1P)$ $1^+$&2427$\pm$36 & 2524&\\
&&$(1P)$ $2^+$&$D^*_{2}(2460)$ & 2588&\\
\end{tabular}
\end{center}
\end{table}

\begin{figure}
\caption{Effective scale-dependent strong coupling constant $\alpha_s$ given
in Eq. (\protect\ref{asf}) as a function of momentum. 
We plot by the solid line our parametrization. 
Dots and triangles are the experimental results
of Refs. \protect\cite{klut} and \protect\cite{emem}, respectively.
For comparison we plot by a dashed line the parametrization obtained
in Ref. \protect\cite{shir} using $\Lambda=$ 0.2 GeV.}
\label{fig1}
\end{figure}

\begin{figure}
\caption{Masses of the $a_J$ triplet members as a function of the
scalar/vector rate confinement, $a_s$. The solid line denotes the 
$a_0$ mass, the dashed that of $a_1$ and the dashed-dotted stands
for the mass of the $a_2$ meson. A vertical solid line indicates
the value chosen for $a_s$.}
\label{fig2}
\end{figure}


\begin{references}

\bibitem{eich} E. Eichten {\it et al.},
		Phys. Rev. Lett. {\bf 34}, 369 (1975).

\bibitem{godf} S. Godfrey and N. Isgur, 
		Phys. Rev. D {\bf 32}, 189 (1985).

\bibitem{manh} A. Manohar and H. Georgi, 
		Nucl. Phys. B {\bf 234}, 189 (1984).

\bibitem{bali} G.S. Bali, 
		Phys. Rep. {\bf 343}, 1 (2001).

\bibitem{lui} L.A. Blanco, F. Fern\'andez, and A. Valcarce, 
		Phys. Rev. C {\bf 59}, 428 (1999);
		R. Bonnaz, L.A. Blanco, B. Silvestre-Brac, F. Fern\'andez,
		and A. Valcarce, 
		Nucl. Phys. A {\bf 683}, 425 (2001).

\bibitem{diak} D.I. Diakonov and V.Yu. Petrov, 
		Nucl. Phys. B {\bf 245}, 259 (1984); 
		{\it ibid} {\bf 272}, 457 (1986).

\bibitem{mosz} S. Moszkowski,
		{\it Nuclear Physics} (New York: Gordon and Breach, 1968) p. 3.

\bibitem{mach} R. Machleidt, K. Holinde, and Ch. Elster,
		Phys. Rep. {\bf 149}, 1 (1987).

\bibitem{scad} M.D. Scadron, 
		Phys. Rev. D {\bf 26}, 239 (1982).

\bibitem{toof} G. 't Hooft,
		Phys. Rev. D {\bf 14}, 3432 (1976).

\bibitem{resa} C.R. M\"unz, J. Resag, B.C. Metsch, and H.R. Petry,
		Nucl. Phys. A {\bf 578}, 418 (1994).

\bibitem{shur} T. Sch\"afer and E.V. Shuryak,
		Rev. Mod. Phys. {\bf 70}, 323 (1998).

\bibitem{bonnm} D. Merten, R. Ricken, M. Koll, B.C. Metsch, and H.R. Petry,
		Eur. Phys. J. A {\bf 13}, 477 (2002).

\bibitem{ruju} A. de R\'{u}jula, H. Georgi, and S.L. Glashow,
		Phys. Rev. D {\bf 12}, 147 (1975).

\bibitem{BHA80} R.K. Bhaduri, L.E. Cohler, and Y. Nogami,
		Phys. Rev. Lett. {\bf 44}, 1369 (1980).

\bibitem{YYYY} J. Weinstein and N. Isgur,
		Phys. Rev. D {\bf 27}, 588 (1983).

\bibitem{tita} S. Titard and F.J. Yndurain,
		Phys. Rev. D {\bf 51}, 6348 (1995);
               A.M. Badalian and V.L. Morgunov,
		{\it ibid} {\bf 60}, 116008 (1999);
               A.M. Badalian and B.L.G. Bakker,
		{\it ibid} {\bf 62}, 094031 (2000).

\bibitem{shir} D.V. Shirkov and I.L. Solovtsov,
		Phys. Rev. Lett. {\bf 79}, 1209 (1997).

\bibitem{bada2} A.M. Badalian and Yu.A. Simonov,
		Yad. Fiz. {\bf 60}, 714 (1997).

\bibitem{pre2} A.C. Mattingly and P.M. Stevenson,
		Phys. Rev. D {\bf 49}, 437 (1994);
               A.M. Badalian and D.S. Kuzmenko, 
		Phys. Rev. D {\bf 65}, 016004 (2001).

\bibitem{halz} F. Halzen, C. Olson, M.G. Olsson and M.L. Stong,
		Phys. Rev. D {\bf 47}, 3013 (1993).

\bibitem{brun} B. Juli\'a-D\'{\i}az, J. Haidenbauer, A. Valcarce,
		and F. Fern\'andez, 
		Phys. Rev. C {\bf 65}, 034001 (2002);
		H. Garcilazo, A. Valcarce, and F. Fern\'andez, 
		Phys. Rev. C {\bf 64}, 058201 (2001);
                D.R. Entem, F. Fern\'andez, and A. Valcarce, 
		Phys. Rev. C {\bf 62}, 034002 (2000).

\bibitem{pre1} C.T.H. Davies {\it et al.},
		Phys. Rev. D {\bf 56}, 2755 (1997).

\bibitem{klut} S. Kluth, hep-ex/0309070.	
	 
\bibitem{emem} L3 Collaboration, P. Achard {\it et al.},
		Phys. Lett. B {\bf 536}, 217 (2002).

\bibitem{weis} W. Weise, 
		Int. Rev. Nucl. Phys. {\bf 1}, 58 (1984), p. 137.

\bibitem{thom} A.W. Thomas, 
		Adv. Nucl. Phys. {\bf 13}, 1 (1984), p. 55.

\bibitem{ferp} F. Fern\'andez, A. Valcarce, P. Gonz\'alez, and V. Vento,
		Phys. Lett. B {\bf 287}, 35 (1992).

\bibitem{yazk} K. Yazaki, 
		Prog. Part. Nucl. Phys. {\bf 24}, 353 (1990).

\bibitem{bali2} G.S. Bali {\it et al.}, 
		Phys. Rev. D {\bf 62}, 054503 (2000).

\bibitem{este} P.W. Stephenson, 
		Nucl. Phys. B {\bf 550}, 427 (1999).
		F. Knechtli and R. Sommer, 
		Nucl. Phys. B {\bf 590}, 309 (2000). 
		O. Philipsen and H. Wittig, 
		Phys. Lett. B {\bf 451}, 146 (1999). 
		P. de Forcrand and O. Philipsen, 
		Phys. Lett. B {\bf 475}, 280 (2000).

\bibitem{miss} J. Vijande, P. Gonz\'alez, H. Garcilazo and A. Valcarce,
		Phys. Rev. D {\bf 69}, 074019 (2004).

\bibitem{golr} M.M. Brisudov\'a, L. Burakovsky, and T. Goldman,
		Phys. Rev. D {\bf 61}, 054013 (2000).

\bibitem{luca} W. Lucha, F.F. Sch\"{o}berl, and D. Gromes, 
		Phys. Rep. {\bf 200}, 127 (1991).

\bibitem{szcz} A.P. Szczepaniak and E.S. Swanson, 
		Phys. Rev. D {\bf 55}, 3987 (1997).

\bibitem{suga} J. Sugiyama, S. Mashita, M. Ishida, and M. Oka,
		hep-ph/0306111.

\bibitem{bram} N. Brambilla and A. Vairo, 
		Nucl. Phys. B (Proc. Supp.) {\bf 64}, 418 
		(1998), and references therein.

\bibitem{bbal} G.S. Bali, A. Wachter, and K. Schilling, 
		Phys. Rev. D {\bf 56}, 2556 (1997), and references therein.

\bibitem{isg2} N. Isgur, 
		Phys. Rev. D {\bf 62}, 014025 (2000); nucl-th/0007008.

\bibitem{bumu} J. Burger, R. M\"uller, K. Tragl, and H.M. Hofmann,
		Nucl. Phys. A {\bf 493}, 427 (1989). 

\bibitem{okann} M. Oka and S. Takeuchi,
		Phys. Rev. Lett. {\bf 63}, 1780 (1989).

\bibitem{shuro} E.V. Shuryak and J.L. Rosner,
		Phys. Lett. B {\bf 218}, 72 (1989).
		 
\bibitem{bonnz} W.H. Blask, U. Bohn, M.G. Huber, B.C. Metsch, and H.R. Petry, 
		Z. Phys. A {\bf 337}, 327 (1990).

\bibitem{bonnep} M. Koll, R. Ricken, D. Merten, B.C. Metsch, and H.R. Petry,
		Eur. Phys. J. A {\bf 9}, 73 (2000).

%\bibitem{pdgb} K. Hagiwara {\it et al.},
%		Phys. Rev. D {\bf 66}, 010001 (2002).
%
\bibitem{pdgb} S. Eidelman {\it et al.},
		Phys. Lett. B {\bf 592}, 1 (2004).

\bibitem{pedr} P. Gonz\'alez, A. Valcarce, H. Garcilazo, and J. Vijande,
       	Phys. Rev. D {\bf 68}, 034007 (2003), and references therein. 

\bibitem{eic1} E.J. Eichten, K. Lane, and C. Quigg,
       	Phys. Rev. D {\bf 69}, 094019 (2004).

\bibitem{lucp} W. Lucha, F.F. Schr\"oberl, and D. Gromes,
		Phys. Rep. {\bf 200}, 127 (1991).

\bibitem{sema} C. Semay and B. Silvestre-Brac, 
		Phys. Rev. D {\bf 46}, 5177 (1992), and references therein.

\bibitem{koon} S.E. Koonin and D.C. Meredith, 
		{\it Computational Physics}, (Addison-Wesley, New York, 1990).

\bibitem{kloe} KLOE Collaboration, A. Aloisio {\it et al.},
		Phys. Lett. B {\bf 541}, 45 (2002).

\bibitem{cbcb} Crystal Barrel Collaboration, C. Amsler {\it et al.}, 
		Phys. Lett. B {\bf 294}, 451 (1992).

\bibitem{gilm} F.J. Gilman and R. Kauffman,
		Phys. Rev. D {\bf 36}, 2761 (1987).

\bibitem{feld} Th. Feldmann, P. Kroll, and B. Stech,
		Phys. Rev. D {\bf 58}, 114006 (1998);
		Phys. Lett. B {\bf 449}, 339 (1999).

\bibitem{brae} A. Bramon, R. Escribano, and M.D. Scadron,
		Eur. Phys. J. C {\bf 7}, 271 (1999).

\bibitem{bura} L. Burakovsky and T. Goldman,
	    Phys. Lett. B {\bf 427},361 (1998).

\bibitem{focu} FOCUS Collaboration, F.L. Fabri {\it et al.},
         	hep-ex/0011044.     

\bibitem{foc2} FOCUS Collaboration, talk presented by R. K. Kutschke
		at the 5th Workshop on Heavy Flavours at Fixed Target,
		Rio de Janeiro, Brazil (2000).

\bibitem{bel5} Belle Collaboration, K. Abe {\it et al.}, 
		Phys. Rev. D {\bf 69}, 112002 (2004).

\bibitem{cleo4} CLEO Collaboration, S. Anderson {\it et al.},
		Nucl. Phys. A {\bf 663}, 647 (2000).

\bibitem{delp} DELPHI Collaboration, talk presented by D. Bloch at the 
		EPS International Conference on High Energy Physics, 
		Budapest, Hungary (2001).

\bibitem{baba} BaBar Collaboration, B. Aubert {\it et al.}, 
		Phys. Rev. Lett. {\bf 90}, 242001 (2003).

\bibitem{bab2} BaBar Collaboration, B. Aubert {\it et al.}, 
		Phys. Rev. D {\bf 69}, 031101 (2004).

\bibitem{cleo} CLEO Collaboration, D. Besson {\it et al.}, 
		Phys. Rev. D {\bf 68}, 032002 (2003).

\bibitem{cleo2} CLEO Collaboration, talk presented by H. Vogel at the 
		International Europhysics Conference on High Energy 
		Physics, Aachen, Germany (2003).

\bibitem{cleo3} CLEO Collaboration, S. Stone and J. Urheim,
		hep-ph/0308166.

\bibitem{bel3} Belle Collaboration, Y. Mikami {\it et al.},
		Phys. Rev. Lett. {\bf 92}, 012002 (2004).

\bibitem{bel4} Belle Collaboration, P. Krokovny {\it et al.}, 
		Phys. Rev. Lett. {\bf 91}, 262002 (2003).

\bibitem{clen} CLEO Collaboration, S.E. Csorna {\it et al.}, hep-ex/0207060.
 
\bibitem{kapa} E791 Collaboration, E.M. Aitala {\it et al.},
         Phys. Rev. Lett. {\bf 89}, 121801 (2002).      

\bibitem{PDG02b} K. Hagiwara {\it et al.},
		Phys. Rev. D {\bf 66}, 010001 (2002).

\bibitem{e852} E852 Collaboration, G.S. Adams {\it et al.},
		Phys. Lett. B {\bf 516}, 264 (2001).

\bibitem{ANIS} V.V. Anisovich,
		AIP Conf. Proc. {\bf 717}, 441 (2004). 

\bibitem{brna} T. Barnes, N. Black, and P.R. Page, 
		Phys. Rev. D {\bf 68}, 054014 (2003).

\bibitem{napo} S. Godfrey and J. Napolitano, 
		Rev. Mod. Phys. {\bf 71}, 1411 (1999).


\bibitem{Ac00m} L3 Collaboration, M. Acciarri {\it et al},
		Phys. Lett. B {\bf 501}, 1 (2001).

\bibitem{LY03} D.M. Li, H. Yu and S.S. Fang,
        Eur. Phys. J. C {\bf 28}, 335 (2003).

\bibitem{bar1} WA102 Collaboration, B. Barberis {\it et al.},
		Phys. Lett. B {\bf 413}, 225 (1997); {\it ibid} 
		{\bf 440}, 225 (1998).

\bibitem{sosa} E690 Collaboration, M. Sosa {\it et al.},
		Phys. Rev. Lett. {\bf 83}, 913 (1999).

\bibitem{clos} F.E. Close and A. Kirk,
		Z. Phys. C {\bf 76}, 469 (1997).

%\bibitem{pdge} Accessible under {\it http://pdg.lbl.gov}.

\bibitem{acha} M.N. Achasov {\it et al.},
		Phys. Rev. D {\bf 66}, 032001 (2002).

\bibitem{link} FOCUS Collaboration, J.M. Link {\it et al.},
		Phys. Lett. B {\bf 545}, 50 (2002).

\bibitem{vlad} V.V. Vladimirsky {\it et al.}, 
		Phys. Atom. Nucl. {\bf 64}, 1895 (2001).

\bibitem{morn} C.J. Morningstar and M. Peardon,
		 Phys. Rev. D {\bf 60}, 034509 (1999).

\bibitem{ache} AFS Collaboration, T. Akesson {\it et al.},
		Nucl. Phys. B {\bf 264}, 154 (1986).

\bibitem{h113} A. Abele {\it et al.},
		Phys. Lett. B {\bf 415}, 280 (1997).

\bibitem{ABEL} Crystal Barrel Collaboration, A. Abele {\it et al.},
		Phys. Lett. B {\bf 391}, 191 (1997).

\bibitem{rosn} J. Rosner, 
		Com. Nucl. Part. Phys. {\bf 16}, 109 (1986).

\bibitem{iswi} N. Isgur and M.B. Wise, 
		Phys. Lett. B {\bf 232}, 113 (1989).

\bibitem{koko} S. Godfrey and R. Kokoski,
		Phys. Rev. D {\bf 43}, 1679 (1991).

\bibitem{cleo5} CLEO Collaboration, D. Besson {\it et al.}, hep-ex/0305017.

\bibitem{luch} W. Lucha and F.F. Sch\"oberl, 
		Mod. Phys. Lett. A {\bf 18}, 2837 (2003), 
		and references therein.

\bibitem{JAFF} R.J. Jaffe, 
		Phys. Rev. D {\bf 15}, 267 (1977).

\bibitem{dean} A. Deandrea, G. Nardulli, and A.D. Polosa, 
        Phys. Rev. D {\bf 68}, 097501 (2003).

\bibitem{bard} W.A. Bardeen, E.J. Eichten, and C.T. Hill,
    	Phys. Rev. D {\bf 68}, 054024 (2003).

\bibitem{bel1} Belle Collaboration, S.-K. Choi {\it et al.}, 
		Phys. Rev. Lett. {\bf 89}, 102001 (2002).

\bibitem{eber} D. Ebert, R.N. Faustov, and V.O. Galkin,
                Phys. Rev. D {\bf 62}, 034014 (2000).

\bibitem{hfs2} E.J. Eichten and C. Quigg,
		Phys. Rev. D {\bf 49}, 5845 (1994).

\bibitem{hfs3} R. Barbieri, R. Gatto, R. K\"ogerler, and Z. Kunszt,
		Phys. Lett. B {\bf 57}, 455 (1975).

\bibitem{chi1} Belle Collaboration, S.-K.Choi {\it et al.},
	    Phys. Rev. Lett. {\bf 91}, 262001 (2003).   
 
\bibitem{chi2} CDF Collaboration, talk presented by G. Bauer at the 
        Second International Workshop on Heavy Quarkonium, Fermilab, USA
		(2003).

\bibitem{chi3} T. Barnes and S. Godfrey, Phys. Rev. D {\bf 69}, 054008 (2004). 

\bibitem{mes4} J. Vijande, talk presented at the MESON2004 Conference, Cracow,
		Poland (2004).

\bibitem{ZZZZ} P. Geiger and N. Isgur, 
		Phys. Rev. Lett. {\bf 67}, 1066 (1991);
		F.E. Close and N.A. T\"ornqvist,
		J. Phys. G {\bf 28}, R249 (2002).

\bibitem{XXXX} N.N. Achasov,
		Nucl. Phys. A {\bf 675}, 279c (2000).

\bibitem{amsl} C. Amsler, 
		Phys. Lett. B {\bf 541}, 22 (2002).

\bibitem{tera} K. Terasaki, 
		AIP Conf. Proc. {\bf 717}, 556 (2004). 

\bibitem{vij2} J. Vijande, F. Fern\'andez and A. Valcarce,
		AIP Conf. Proc. {\bf 717}, 352 (2004). 

\bibitem{blac} D. Black, A.H. Fariborz, and J. Schechter,
		Phys. Rev. D {\bf 61}, 074001 (2000).

\bibitem{AMS2} C. Amsler and F.E. Close,
		Phys. Rev. D {\bf 53}, 295 (1996). 

\bibitem{WEIN} W. Lee and D. Weingarten,
		Phys. Rev. D {\bf 61}, 014015 (2000) and references therein.

\bibitem{ani2} A.V. Anisovich {\it et al.}, 
		Nucl. Phys. A {\bf 662}, 319 (2000).

\bibitem{umek}  T. Umekawa,
		hep-ph/0306040.

\bibitem{PRD52} G. Janssen, B.C. Pearce, K. Holinde and J. Speth,
		Phys. Rev. D {\bf 52}, 2690 (1995). 

\bibitem{TORN} N.A. Tornqvist, 
         Z. Phys. C {\bf 68}, 647 (1995).

\bibitem{PENI} M. Boglione and M.R. Pennington, 
         Phys Rev Lett {\bf 79}, 1998 (1997).

\bibitem{bacl} T. Barnes, F.E. Close, and H.J. Lipkin,
		Phys. Rev. D {\bf 68}, 054006 (2003).

\bibitem{vij4} J. Vijande, F. Fern\'andez, A. Valcarce, and B. Silvestre-Brac,
		Eur. Phys. J. A {\bf 19}, 383 (2004).

\bibitem{futu} J. Vijande, F. Fern\'andez, and A. Valcarce, 
		work in progress.

\end{references}
\end{document}